\documentclass[11pt,paper]{JHEP3}

\usepackage{amsmath,amssymb,array,calc,epsfig}

\newlength{\intwidth}

\title{Taming Tree Amplitudes In General Relativity}

\author{Paolo Benincasa\\
Department of Applied Mathematics, University of
Western Ontario, London, Ontario N6A 5B7, Canada\\
Email: \email{pbeninca@uwo.ca}}
\author{Camille Boucher-Veronneau\\
Department of Physics and Astronomy, University of
Waterloo, Waterloo, Ontario N2L 3G1, Canada\\
Perimeter Institute for Theoretical Physics, Waterloo,
Ontario N2J 2W9, Canada\\
Email: \email{cboucher-veronneau@perimeterinstitute.ca}}
\author{Freddy Cachazo\\
Perimeter Institute for Theoretical Physics, Waterloo,
Ontario N2J 2W9, Canada\\
Email: \email{fcachazo@perimeterinstitute.ca}}

\abstract{
We give a proof of BCFW recursion relations for all tree-level
amplitudes of gravitons in General Relativity. The proof follows the
same basic steps as in the BCFW construction and it is an extension
of the one given for next-to-MHV amplitudes by one of the authors
and P. Svr\v{c}ek in hep-th/0502160. The main obstacle to overcome
is to prove that deformed graviton amplitudes vanish as the complex
variable parameterizing the deformation is taken to infinity. This
step is done by first proving an auxiliary recursion relation where
the vanishing at infinity follows directly from a Feynman diagram
analysis. The auxiliary recursion relation gives rise to a
representation of gravity amplitudes where the vanishing under the
BCFW deformation can be directly proven. Since all our steps are
based only on Feynman diagrams, our proof completely establishes the
validity of BCFW recursion relations. This means that results
in the literature that were derived assuming their validity become
true statements.
}


\preprint{UWO-TH-07/04}

\begin{document}

\section{Introduction}

The analytic computation of scattering amplitudes in gauge theory
and gravity has always been a very challenging problem. In
principle, this problem is solved by using Feynman diagrams.
However, in practice, the fast growth in the number of diagrams
makes the calculation impossible.

In some cases, closed formulas have been found for large classes of
amplitudes. One of the main tools has been the use of recursion
techniques. Many analytic formulas were found or proven by using the
Berends-Giele recursion relations introduced in the
80's~\cite{Berends:1987me,Mangano:1987xk,Mangano:1990by,Berends:1989hf,Kosower:1989xy}.
One important example are the wonderfully simple formulas
conjectured by Parke-Taylor~\cite{Parke:1986gb} for MHV (Maximally
Helicity Violating) tree level amplitudes of gluons.

More recently, a new set of recursion relations for tree level
amplitudes of gluons was introduced by Britto, Feng and the third
author~\cite{Britto:2004ap}. These recursion relations were inspired
by \cite{Britto:2004nc,Roiban:2004ix} and reproduced very compact
results obtained in \cite{Bern:2004ky} by studying the IR behavior
of ${\cal N}=4$ one-loop amplitudes. A simple and elegant proof of
the relations was later given by the same authors in collaboration
with Witten in \cite{Britto:2005fq}. The proof is constructive and
gives rise to a method using the power of complex analysis for
deriving similar relations in any theory where physical
singularities are well understood. The BCFW method has been
successfully applied in many contexts involving massless particles
at tree and loop level
\cite{Luo:2005rx,Luo:2005my,Bern:2005ji,Bern:2005cq,
Berger:2006ci,Berger:2006vq,Bern:2005hh,Brandhuber:2007up} as well
as for massive particles at tree level \cite{Badger:2005zh}.

The possibility of the existence of BCFW recursion relations in
General Relativity was first investigated in
\cite{Bedford:2005yy,Cachazo:2005ca}. There it was pointed out that
the main obstacle to establish the validity of the recursion
relations is to prove that deformed amplitudes vanish at infinity
while individual Feynman diagrams diverge. In \cite{Bedford:2005yy},
the desired behavior was checked for MHV amplitudes up to $n<11$
under the $(-,-)$ deformation using the BGK formula
\cite{Berends:1988zp}. In \cite{Cachazo:2005ca}, it was shown that
the BGK formula vanishes at infinity for any $n$\footnote{Since the
BGK formula has been tested against Feynman diagrams only for $n<11$
 \cite{Berends:1988zp}, one cannot make a general statement for
actual amplitudes based on BGK.} under the $(+,-)$ deformation. Also
in \cite{Cachazo:2005ca}, a proof based on Feynman diagrams was
given for all next-to-MHV amplitudes\footnote{Although not mentioned
in \cite{Cachazo:2005ca}, this technique clearly also works for MHV
amplitudes.} and for all amplitudes up to eight gravitons using the
KLT relations \cite{Kawai:1985xq}.

The fact that individual Feynman diagrams diverge very badly in the
limit when the deformation parameter is taken to infinity and yet
the amplitude vanishes implies that a large number of cancelations
must happen. What was shown in \cite{Cachazo:2005ca} is that such
cancelations can be made explicit if representations of amplitudes
where Feynman diagrams have been re-summed are used.

This is just one more example where Feynman diagrams not only give
rise to extremely long answers which then collapse to very compact
expressions but actually imply a completely wrong behavior of the
amplitude for large momenta.

A surprising example of this, now at the loop level, is the work of
\cite{Green:2006gt,Bern:2006kd,Green:2006yu} where ${\cal N}=8$
supergravity has been shown to possess a remarkably good ultraviolet
behavior even though a direct power counting argument indicates that
bad divergencies must be present. Also recently, a careful study of
the structure of certain one-loop amplitudes in ${\cal N}=8$
supergravity shows that even though power counting implies that
after a Passarino-Veltman reduction \cite{Passarino:1978jh} the
amplitude should contain boxes, triangles, bubbles and rational
pieces only the boxes can have non-zero coefficients
\cite{Bern:2005bb,Bjerrum-Bohr:2006yw}. That this might hold for
generic one-loop amplitudes is now known as the no-triangle
hypothesis \cite{Bern:2005bb,Bjerrum-Bohr:2006yw}. A striking
possibility, which could explain all these properties, is that a
twistor string-like construction for this theory could exists
\cite{Witten:2003nn,Nair:2005iv,Abou-Zeid:2006wu}.

In this paper we give a complete proof that the miraculous behavior
exhibited in next-to-MHV tree level amplitudes of gravitons in
\cite{Cachazo:2005ca} actually extends to all amplitudes.

The strategy we follow is exactly the same as the one used in
\cite{Cachazo:2005ca} to prove the next-to-MHV case. We use an
auxiliary recursion relation to derive a more convenient
representation for the amplitudes and then show that they vanish at
infinity under the BCFW deformation.

The most important aspect of our proof is that both the auxiliary
recursion relations and the vanishing under the BCFW deformation are
proven using only Feynman diagram arguments. Since Feynman diagrams
are the basic way to define gravity amplitudes, our result
completely establishes the validity of the BCFW recursion relations
for General Relativity.

The auxiliary recursion relations are obtained by using a
deformation that affects the maximum possible number of polarization
tensors while keeping propagators linear functions in the
deformation parameter. Such a deformation was also introduced in
\cite{Cachazo:2005ca}. Quite interestingly, this ``maximal"
deformation on a given amplitude induces non-maximal deformations on
amplitudes with smaller number of gravitons. One of the non-maximal
deformations that naturally shows up only affects gravitons of a
given helicity. Very interesting results have been obtained in the
literature by assuming that under such deformations amplitudes
vanish at infinity. More precisely, Bjerrum-Bohr et.al were able to
derive MHV expansions for gravity in \cite{Bjerrum-Bohr:2005jr}
along the same lines as done for gauge theory by Risager in
\cite{Risager:2005vk}. As mentioned in \cite{Bjerrum-Bohr:2005jr}, a proof of the validity of the
basic BCFW deformation will constitute evidence for the validity of the non-maximal
deformations as some of them can be thought of as compositions of the
basic one\footnote{After the first version of this paper was submitted, \cite{Bianchi:2008pu} found by direct numerical
analysis that non-maximal deformations in gravity generically fail for
$n\ge12$. This means that the MHV expansion for gravity cannot be obtained in the way proposed in \cite{Bjerrum-Bohr:2005jr}. See the note added for more details and appendix B for a proof.}.

It is also important to mention that at one-loop in gauge theory one
can find that compositions of BCFW deformations can vanish at
infinity while individual deformations do not. This was actually the
motivation for the first use of compositions in the literature in
\cite{Bern:2005hs}.

This paper is organized as follows: In section II, we follow the
same steps as in the original BCFW construction to show the form of
recursion relations for gravity amplitudes that can be obtained if
one assumes that the amplitudes vanish at infinity under the
deformation. In section III we prove that statement by using
auxiliary recursion relations. In section IV, we use Ward identities
for MHV amplitudes to show how our proof implies the validity of
other recursion relations obtained by different deformations. In
section V we give our conclusions and future directions. Part of the
proof of the validity of the auxiliary recursion relations is given
in the appendix.

\subsection{Preliminaries And Conventions}

Tree level amplitudes of gravitons are rational functions of the
momenta of the gravitons and multilinear functions of the
polarization tensors. It is convenient to encode all the information
in terms of spinor variables using the spinor-helicity formalism
\cite{Berends:1981rb,DeCausmaecker:1981bg,Kleiss:1985yh}. Each
momentum vector can be written as a bispinor $p_{a\dot a} =
\lambda_a\tilde{\lambda}_{\dot a}$. We define the inner product of
spinors as follows $\langle \lambda ,\lambda'\rangle =
\epsilon^{ab}\lambda_a\lambda'_b$ and $[ \tilde\lambda
,\tilde\lambda'] = \epsilon^{\dot a \dot b}\tilde\lambda_{\dot
a}\tilde\lambda'_{\dot b}$. Polarization tensors of gravitons can be
expressed in terms of polarization vectors of gauge bosons as follow
\begin{equation}
\epsilon^+_{a\dot a, b\dot b} = \epsilon_{a\dot a}^+\epsilon_{b\dot
b}^+ \, , \qquad \epsilon^-_{a\dot a, b\dot b} = \epsilon_{a\dot
a}^-\epsilon_{b\dot b}^-
\end{equation}
where polarization vectors of gauge bosons are given by
\begin{equation}
\epsilon^+_{a\dot a} = \frac{\mu_a\tilde\lambda_{\dot a}}{\langle
\mu ,\lambda \rangle}, \qquad \epsilon^-_{a\dot a}
=\frac{\lambda_a\tilde\mu_{\dot a}}{[\tilde\lambda , \tilde\mu]}
\end{equation}
with $\mu_a$ and $\tilde\mu_{\dot a}$ arbitrary reference spinors.

Using the spinor-helicity formalism all the information about a
particular graviton is encoded in $\lambda,\tilde\lambda$ and the
helicity, $h$, which can be positive or negative. Therefore a given
amplitude can be written as
\begin{equation}
{\cal M}_n(1^{h_1},\ldots , n^{h_n}) =\kappa^{n-2}
\delta^{(4)}\left(\sum_{i=1}^n\lambda^{(i)}_a\tilde\lambda^{(i)}_{\dot
a}\right)M_n(1^{h_1},\ldots , n^{h_n}),
\end{equation}
where $\kappa^2 = 8\pi G_{\rm N}$, the label $(i)$ on the spinors is
the particle label and the notation $(i^{h_i})$ stands for
$(\lambda^{(i)}, \tilde\lambda^{(i)}, h_i)$. In the rest of this
paper we will only be concerned with $M_n(1^{h_1},\ldots ,
n^{h_n})$.

Sometimes it will be convenient to write $(i^{h_i})$ as $p_i^{h_i}$
where $p_i$ is the momentum of the $i^{\rm th}$ graviton.

Also useful is the following notation: $\langle \lambda | P |
\tilde\lambda'] = -\lambda^a P_{a\dot a}\tilde\lambda^{\dot a}$. The
minus sign in the definition is there so that if $P_{a\dot a}$ is a
null vector $\mu_a\tilde\mu_{\dot a}$ one has $\langle \lambda | P |
\tilde\lambda'] = \langle \lambda, \mu\rangle
[\tilde\mu,\tilde\lambda]$. This formula has several
generalizations. In this paper we only use the one that involves two
generic vectors $P$ and $Q$ that are written as sums of null vectors
as $P=\sum_{s}p_s$ and $Q=\sum_{r}p_r$. Then we have
\begin{equation}
\langle \lambda | P \; Q | \lambda' \rangle = \sum_{r,s}\langle
\lambda, \lambda^{(r)} \rangle
[\tilde\lambda^{(r)},\tilde\lambda^{(s)}] \langle \lambda^{(s)},
\lambda' \rangle.
\end{equation}

\section{BCFW Construction For Gravity Amplitudes}

Consider a scattering amplitude of $n$ gravitons $M_n(1^{h_1},\ldots
, n^{h_n})$. Construct a one complex parameter deformation of the
amplitude that preserves the physical properties of being on-shell
and momentum conservation. The simplest way to achieve this is by
choosing two gravitons of opposite helicities\footnote{This is
always possible since tree-level amplitudes with all equal
helicities vanish and are not of interest for our discussion.}, say
$i^+$ and $j^-$, and perform the following deformation
\begin{equation}
\label{bcfw} \lambda^{(i)}(z) = \lambda^{(i)} + z\lambda^{(j)},
\qquad \tilde\lambda^{(j)}(z) = \tilde\lambda^{(j)} -
z\tilde\lambda^{(i)}.
\end{equation}
All other spinors remain the same. The deformation parameter $z$ is
a complex variable. It is easy to check that this deformation
preserves the on-shell conditions of all gravitons, i.e., $p_k(z)^2
= 0$ for any $k$ and momentum conservation since $p_i(z)+p_j(z) =
p_i+p_j$.

The main observation is that the scattering amplitude is a rational
function of $z$ which we denote by $M_n(z)$. This fact follows from
$M_n(1^{h_1},\ldots , n^{h_n})$ being a rational function of momenta
and polarization tensors. Being a rational function of $z$, $M_n(z)$
can be determined if complete knowledge of its poles, residues and
behavior at infinity is found.

We claim that $M_n(z)$ only has simple poles and it vanishes as $z$
is taken to infinity. This means that
\begin{equation}
M_n(z) = \sum_{\alpha}\frac{c_\alpha}{z-z_\alpha}
\end{equation}
where the sum is over all poles of $M_n(z)$.

The fact that $M_n(z)$ only has simple poles follows by considering
its form as a sum over Feynman diagrams. Choosing a gauge where
polarization tensors do not have poles in $z$, i.e, one in which the
reference spinors of the $i^{\rm th}$ and $j^{\rm th}$ gravitons are
$\mu_a = \lambda^{(j)}_{a}$ and $\tilde\mu_{\dot a} =
\tilde\lambda_{\dot a}^{(i)}$ respectively, the only possible
singularities come from propagators. Propagators are functions of
momenta of the form
\begin{equation}
\frac{1}{P^2_{\cal I}} = \frac{1}{(\sum_{k\in \cal I}p_k)^2}
\end{equation}
where ${\cal I} \subset \{ 1,2,\ldots, n\}$ is some subset of
gravitons with more than one and less than $n-1$ elements.

Clearly, the only propagators that can depend on $z$ are those for
which either $i\in\cal I$ or $j\in \cal I$ but not both. Without
loss of generality let us assume that $i\in \cal I$. Then the
propagator has the form
\begin{equation}
\frac{1}{P^2_{\cal I}(z)} = \frac{1}{P^2_{\cal I}(0) -z\langle
j|P_{\cal I}(0)|i]}.
\end{equation}
This shows that all singularities are simple poles. Their location
is given by
\begin{equation}
z_{\cal I} = \frac{P^2_{\cal I}(0)}{\langle j|P_{\cal I}(0)|i]}.
\end{equation}

The proof that $M_n(z)$ vanishes as $z$ is taken to infinity is
basically the main result of this paper and it is presented in the
next section. Here we simply assume it and continue in order to
present the final form the BCFW recursion relations.

The final step is the computation of the residues $c_{\cal I}$. This
is easily done since close to the region where a given propagator
goes on-shell the amplitude factorizes as the product of lower
amplitudes. Collecting all these results one finds that
\begin{equation}
\label{recu} M_n(z) = \sum_{{\cal I},{\cal J}}\sum_{h=\pm} M_{\cal
I}\left( \{K_{\cal I}\}, p_i(z_{\cal I}), -P_{\cal I}^h(z_{\cal I})
\right)\frac{1}{P_{\cal I}(z)^2} M_{\cal J}\left( \{K_{\cal J}\},
p_j(z_{\cal I}), P_{\cal I}^{-h}(z_{\cal I}) \right)
\end{equation}
where $\{ \cal I , \cal J\}$ is a partition of the set of all
gravitons such that $i\in {\cal I}$ and $j\in {\cal J}$, $K_{\cal
I}$ ($K_{\cal J}$) is the collection of all gravitons in $\cal I$
($\cal J$) except for $i$ ($j$) and $h$ is the helicity of the
internal graviton.

The BCFW recursion relation is obtained by setting $z=0$ in
(\ref{recu}). It is important to mention that the value of $z_{\cal
I}$ was determined by requiring $P_{\cal I}(z_{\cal I})$ be a null
vector. Therefore the BCFW recursion relations only involve physical
on-shell amplitudes.

\section{Vanishing Of $M_n(z)$ At Infinity}

In the previous section we showed that the validity of the BCFW
recursion relations for gravity amplitudes simply follows from the
vanishing of $M_n(z)$ at infinity. In this section we provide a
proof of this statement.

It is instructive to start by computing what the behavior of
$M_n(z)$ for large $z$ is from a naive Feynman diagram
analysis\footnote{The reason we use the word ``naive" is that the
argument only takes into account the behavior of individual diagrams
and does not consider possible cancelations among them.}. A generic
Feynman diagram is schematically given by the product of
polarization tensors, propagators and vertices. We are looking for
Feynman diagrams that give the leading behavior for large $z$. We
choose generic reference spinors in polarization tensors such that
\begin{equation}
\epsilon^{++}_i(z)_{a{\dot a},b{\dot b}} \sim \frac{1}{z^2}
\frac{\mu_a\tilde\lambda^{(i)}_{\dot
a}\mu_b\tilde\lambda^{(i)}_{\dot b}}{\langle \mu,
\lambda^{(j)}\rangle^2}, \qquad \epsilon^{--}_j(z)_{a{\dot a},b{\dot
b}}\sim \frac{1}{z^2} \frac{\lambda^{(j)}_a\tilde\mu_{\dot
a}\lambda^{(j)}_b\tilde\mu_{\dot b}}{[\tilde\lambda^{(i)},\tilde\mu
]^2},
\end{equation}
while all others are independent of $z$. Note that only vertices
that depend on momenta can give $z$ contributions in the numerator.
Therefore we should look for Feynman diagrams with the maximum
number of $z$ dependent vertices. Such diagrams are those for which
one has only cubic vertices. For $n$ gravitons there can be a
maximum of $n-2$ vertices. Each vertex can give at most a $z^2$
dependence\footnote{This and all statements about the general
structure of Feynman diagrams can be easily derived from the
lagrangian density ${\cal L} = \sqrt{-g}R$ with $g_{\mu\nu} =
\eta_{\mu\nu}+h_{\mu\nu}$.}. Therefore, the leading diagrams will
have a $z^{2(n-2)}$ dependence from vertices. Finally we are left
with propagators. The $z$ dependence flows in the diagram along a
unique path connecting the $i^{\rm th}$ graviton with the $j^{\rm
th}$ graviton. Therefore there are $n-3$ of them. Each propagator
gives a $1/z$ contribution. Collecting all contributions gives
\begin{equation}
M_n(z)\sim
\left(\frac{1}{z^4}\right)\left(z^{2(n-2)}\right)\left(\frac{1}{z^{(n-3)}}\right)
= z^{n-5}.
\end{equation}
This implies that $M_n(z)\sim 0$ for large $z$ only if $n<5$. As $n$
increases individual Feynman diagrams diverge more at infinity.

This means that we have to find a better representation of $M_n(z)$
where Feynman diagrams have been re-summed into better behaved
objects. This is the main strategy of our proof.

The proof is straightforward but it might be somewhat confusing if
an overall picture is not kept in mind. This is why we first provide
an outline and then give the details.

\subsection{Outline Of The Proof}

We start by finding a convenient representation of $M_n(z)$. The new
representation comes from some auxiliary recursion relations. The
auxiliary recursion relations are obtained using a BCFW-like
construction but with a deformation under which individual Feynman
diagrams vanish at infinity. The way we achieve this is by making as
many polarization tensors go to zero at infinity as possible.

Let us denote the new deformation parameter $w$. Then one has that
$M_n(w)\to 0$ as $w\to\infty$. The recursion relations are
schematically of the form
\begin{equation}
\label{auxily} M_n =\sum_{{\cal I},{\cal J}}\sum_{h=\pm} M_{\cal
I}^h(w_{\cal I})\frac{1}{P^2_{\cal I}}M_{\cal J}^{-h}(w_{\cal I})
\end{equation}
where the sum is over some sets ${\cal I},{\cal J}$ of gravitons.
These auxiliary recursion relations actually provide the first
example of recursion relations valid for all physical amplitudes of
gravitons. However, the price one pays for being able to prove that
$M_n(w)\to 0$ as $w\to\infty$ directly from Feynman diagrams is that
the number of terms in (\ref{auxily}) is very large and many of the
gravitons depend on $w_{\cal I}$. These features make (\ref{auxily})
not very useful for actual computations.

The next step in our proof is to apply the BCFW deformation to $M_n$
now given by (\ref{auxily}). Then we have
\begin{equation}
\label{alik} M_n(z) =\!\!\!\! \sum_{\{i,j\} \subset {\cal J}}\sum_{h
=\pm} M_{\cal I}^h(w_{\cal I})\frac{1}{P^2_{\cal I}}M_{\cal
J}^{-h}(w_{\cal I},z) + \!\!\!\! \sum_{i\in {\cal I}, j\in {\cal
J}}\sum_{h=\pm} M_{\cal I}^{h}(w_{\cal I}(z),z)\frac{1}{P^2_{\cal
I}(z)}M_{\cal J}^{-h}(w_{\cal I}(z),z)
\end{equation}
where the $z$ dependence on the right hand side can appear
implicitly through $w_{\cal I}(z)$ as well as explicitly. The first
set of terms on the right hand side of (\ref{alik}) has both
deformed gravitons in ${\cal J}$. Therefore, all the $z$ dependence
is confined to $M_{\cal J}$. We then show that $M_{\cal J}$ is a
physical amplitude with less than $n$ gravitons under a BCFW
deformation. Therefore, we can use an induction argument to prove
that it vanishes as $z\to \infty$.

For the second set of terms the $z$ dependence appears not only
explicitly but also implicitly via $w_{\cal I}$ in many gravitons.
Quite nicely, it turns out that one can show that each one of those
terms vanishes as $z$ goes to infinity by using a Feynman diagram
analysis similar to the one done at the beginning of this section.
The reason for this is again the large number of polarization
tensors that pick up a $z$ dependence.

There is a special case that has to be considered separately. This
is when there is only one positive helicity graviton in ${\cal I}$,
i.e., the $i^{\rm th}$ graviton. We prove the desired behavior at
infinity in this case at the end of this section.

\subsection{Auxiliary Recursion Relation}

The auxiliary recursion relations we need are obtained by using a
composition of BCFW deformations introduced in \cite{Cachazo:2005ca}
and which was used to prove the vanishing of $M_n(z)$ for
next-to-MHV amplitudes. The basic idea comes form the analysis of
Feynman diagrams we performed above. It is clear that the reason
individual Feynman diagrams diverge as $z\to \infty$ for $n\geq 5$
is that the number of propagators and vertices grow in the same way
but vertices give an extra power of $z$ which can be compensated by
two polarization tensors that depend on $z$ only if $n$ is not too
large. The key is then to perform a deformation that will make more
polarization tensors contribute.

Recall from the outline of the proof that the deformation parameter
is denoted by $w$. The simplest choice is to deform the $\lambda$'s
of all positive helicity gravitons and the $\tilde\lambda$'s of all
negative helicity gravitons. This choice will give $1/w^{2n}$ from
the polarization tensors. This makes $M_n(w)$ go at most as
$1/w^{4}$ even without taking into account the propagators.
Propagators are now quadratic functions of $w$ and therefore they
contribute $1/w^2$ each. This last feature is what makes this choice
very inconvenient since every multi-particle singularity of the
amplitude will result in two simple poles rather than one.

We are then looking for a deformation that gives a $w$ dependence to
the largest number of gravitons and at the same time keeps all
propagators at most linear functions of $w$. The most general such
deformation depends on the number of plus and minus helicity
gravitons in the amplitude. Let $\{r^-\}$ and $\{k^+\}$ denote the
sets of negative and positive helicity gravitons in the amplitude
respectively. Also let $m$ and $p$ be the number of elements in
each. Then if $p\geq m$ the deformation is
\begin{equation}
\label{csdgen} \tilde\lambda^{(j)}(w) = \tilde\lambda^{(j)} - w
\sum_{s\in \{k^+\}}\alpha^{(s)}\tilde\lambda^{(s)}, \qquad
\lambda^{(k)}(w) = \lambda^{(k)} + w \alpha^{(k)}\lambda^{(j)},
\qquad \forall \; k\in \{k^+\}
\end{equation}
where $j$ is a negative helicity graviton and $\alpha^{(k)}$'s can
be arbitrary rational functions of kinematical invariants.

If $m\geq p$ the deformation is
\begin{equation}
\label{csdtgen} \lambda^{(i)}(w) = \lambda^{(i)} + w \sum_{s\in
\{r^-\}}\alpha^{(s)}\lambda^{(s)}, \qquad \tilde\lambda^{(k)}(w) =
\tilde\lambda^{(k)} - w \alpha^{(k)}\tilde\lambda^{(i)}, \qquad
\forall \; k\in \{ r^- \}
\end{equation}
where $i$ is a positive helicity graviton.

The deformation introduced in \cite{Cachazo:2005ca} to prove the
case of next-to-MHV amplitudes corresponds to taking all
$\alpha^{(s)} =1$ in (\ref{csdgen}). It turns out that not all
choices of $\alpha^{(s)}$ lead to the desired behavior of individual
Feynman diagrams at infinity. For example, any choice that removes
the $w$ dependence on any single spinor or even on any linear
combination of subsets of them will fail. This is usually due to
some subtle Feynman diagrams. It is interesting that one has to use
precisely the maximal choice. In other words, we have to choose all
$\alpha^{(s)} = 1$. Given that this is the choice we use in the rest
of the paper, we rewrite (\ref{csdgen}) and (\ref{csdtgen}) with
$\alpha^{(k)}=1$ for later reference.

For $p\geq m$:
\begin{equation}
\label{csd} \tilde\lambda^{(j)}(w) = \tilde\lambda^{(j)} - w
\sum_{s\in \{k^+\}}\tilde\lambda^{(s)}, \qquad \lambda^{(k)}(w) =
\lambda^{(k)} + w \lambda^{(j)}, \qquad \forall \; k\in \{k^+\}
\end{equation}
and $j$ a negative helicity graviton.

If $m\geq p$ the deformation is
\begin{equation}
\label{csdt} \lambda^{(i)}(w) = \lambda^{(i)} + w \sum_{s\in
\{r^-\}}\lambda^{(s)}, \qquad \tilde\lambda^{(k)}(w) =
\tilde\lambda^{(k)} - w \tilde\lambda^{(i)}, \qquad \forall \; k\in
\{ r^- \}
\end{equation}
and $i$ a positive helicity graviton.

The proof that this choice gives $M_n(w)\to 0$ as $w\to \infty$ and
more details are given in the appendix. The proof involves a careful
analysis of when the $w$ can possibly drop out of propagators. This
is basically the point where all other deformations fail.

Here we simply give the final form of the auxiliary recursion
relations. Again we have to distinguish cases. If $p\geq m$ we write
$M_n$ as sums of products of amplitudes with less than $n$ gravitons
as follows:
\begin{equation}\label{RR}
\begin{split}
M_{n}&(\{ r^-\}, \{ k^+\})\:=\\
&=\:\sum_{\mathcal{I}}\sum_{h=\pm}
M_{\mathcal{I}}\left(\left\{r_{\mathcal{I}}^{-}\right\},
 \left\{k_{\mathcal{I}}^{+}(w_{\mathcal{I}})\right\},
 -P_{\mathcal{I}}^{h}(w_{\mathcal{I}})\right)\frac{1}{P^{2}_{\mathcal{I}}}
M_{\mathcal{J}}\left(\left\{r_{\mathcal{J}}^{-}(w_{\cal I})\right\},
 \left\{k_{\mathcal{J}}^{+}(w_{\mathcal{I}})\right\},
 P_{\mathcal{I}}^{-h}(w_{\mathcal{I}})\right)
\end{split}
\end{equation}
where:
\begin{itemize}
\item $\mathcal{I}$ and $\mathcal{J}$ are subsets of the set
      $\left\{1,\ldots,n\right\}$ such that $\mathcal{I}\cup\mathcal{J}\:=\:
      \left\{1,\ldots,n\right\}$. The sum is over all partitions $\{{\cal I},{\cal J}\}$
      of $\{1,\ldots ,n\}$ such that at least one positive helicity graviton is in $\cal I$ and
      $j\in {\cal J}$.
\item $P_{\mathcal{I}}$ is the sum of all the momenta
      of gravitons in $\mathcal{I}$;
\item $\left\{r_{\mathcal{I}}^{-}\right\} \equiv {\cal I}^-$ is the set of negative helicity
      gravitons in $\mathcal{I}$;
\item $\left\{r_{\mathcal{J}}^{-}(w_{\cal I})\right\}$ is the set of negative helicity
      gravitons in $\mathcal{J}$. The $w_{\cal I}$
      dependence is only through $\tilde\lambda^{(j)}(w_{\cal I})$;
\item $\left\{k_{\mathcal{I}}^{+}(w_{\mathcal{I}})\right\} \equiv {\cal I}^+$ is the set of
      positive helicity gravitons in $\mathcal{I}$.
      All of them have been deformed and their dependence on $w_{\cal I}$ is only through
      \begin{equation}\label{spinorz}
       \lambda^{(k)}(w_{\mathcal{I}})\:=\:\lambda^{(k)}+
       w_{\mathcal{I}}\lambda^{(j)};
      \end{equation}
\item $\left\{k_{\mathcal{J}}^{+}(w_{\mathcal{I}})\right\}$ is the set of
      positive helicity gravitons in $\mathcal{J}$.
      All of them have also been deformed via (\ref{spinorz}).
\item The deformation parameter is given by
      \begin{equation}\label{pole}
       w_{\mathcal{I}}\:=\:\frac{P_{\mathcal{I}}^{2}}{
       \sum_{k\in\mathcal{I^+}}\left<j|P_{\mathcal{I}}|k\right]
       }\; .
      \end{equation}
      This definition ensures that the momentum
      \begin{equation}\label{momentumz}
      P_{\mathcal{I}}(w_{\mathcal{I}})_{a\dot
      a}\:=\:P_{\mathcal{I}\: a\dot a}
     +w_{\mathcal{I}}\lambda^{(j)}_a\sum_{k\in\mathcal{I}^+}\tilde\lambda^{(k)}_{\dot a}
      \end{equation}
      is a null vector, i.e., $P_{\cal I}(w_{\cal I})^2 = 0$.
\end{itemize}

Now, if $m\geq p$ then we write $M_n$ as a sum over terms involving
the product of amplitudes with less than $n$ gravitons as follows:
\begin{equation}\label{RRminus}
\begin{split}
M_{n}&(\{ r^-\}, \{ k^+\})\:=\\
&=\:\sum_{\mathcal{I}}\sum_{h=\pm}
M_{\mathcal{I}}\left(\left\{r_{\mathcal{I}}^{-}(w_{\mathcal{I}})\right\},
 \left\{k_{\mathcal{I}}^{+}(w_{\cal I})\right\},
 -P_{\mathcal{I}}^{h}(w_{\mathcal{I}})\right)\frac{1}{P^{2}_{\mathcal{I}}}
M_{\mathcal{J}}\left(\left\{r_{\mathcal{J}}^{-}(w_{\mathcal{I}})\right\},
 \left\{k_{\mathcal{J}}^{+}\right\},
 P_{\mathcal{I}}^{-h}(w_{\mathcal{I}})\right)
\end{split}
\end{equation}
where most definitions are as in the $p\geq m$ case except that the
sets $\cal I$ and $\cal J$ are such that $i\in {\cal I}$ and all the
negative helicity gravitons and the $i^{\rm th}$ positive helicity
graviton are deformed via (\ref{csdt}) instead of (\ref{csd}).

The two rules, (\ref{RR}) and (\ref{RRminus}), provide a full set of
recursion relations for gravity amplitudes. To see this note that
using them one can express any $n$-graviton amplitude as the sum of
products of two amplitudes with less than $n$ gravitons. The smaller
amplitudes which depend on deformed spinors and the intermediate
null vector $P(w_{\cal I})$ are completely ``physical" in the sense
that by construction their momenta are on-shell and satisfy momentum
conservation. Therefore they admit a definition in terms of Feynman
diagrams again and can serve as a starting point to apply either
(\ref{RR}) or (\ref{RRminus}), depending on the new number of plus
and minus helicity gravitons.

\subsection{Induction And Feynman Diagram Argument}

Consider any $n$-graviton amplitude under the BCFW deformation
(\ref{bcfw}) on gravitons $i^+$ and $j^-$:
\begin{equation}
\label{awes} \lambda^{(i)}(z) = \lambda^{(i)} + z\lambda^{(j)},
\qquad \tilde\lambda^{(j)}(z) = \tilde\lambda^{(j)} -
z\tilde\lambda^{(i)}.
\end{equation}

Without loss of generality we can assume that $M_n$ has $p\geq m$
and use (\ref{RR}) as our starting point. If $m\geq p$ we use
(\ref{RRminus}) and everything that follows applies equally well.

Note that the choice of deformed gravitons in (\ref{awes}) is
correlated to that in (\ref{RR}) or (\ref{RRminus}).

Our goal now is to prove that by using (\ref{awes}) on (\ref{RR})
the function $M_n(z)$ vanishes as $z$ is taken to infinity.

Let us consider each term in the sum of (\ref{RR}) individually.
There are two classes of terms. The first kind is when
$\{i,j\}\subset {\cal J}$. The second kind is when $i\in {\cal I}$
and $j\in {\cal J}$.

Consider a term of the first kind,
\begin{equation}
\label{fiki} \sum_{h=\pm}
M_{\mathcal{I}}\left(\left\{r_{\mathcal{I}}^{-}\right\},
 \left\{k_{\mathcal{I}}^{+}(w_{\mathcal{I}})\right\},
 -P_{\mathcal{I}}^{h}(w_{\mathcal{I}})\right)\frac{1}{P^{2}_{\mathcal{I}}}
M_{\mathcal{J}}\left(\left\{r_{\mathcal{J}}^{-}(w_{\cal I},
z)\right\},
 \left\{k_{\mathcal{J}}^{+}(w_{\mathcal{I}},z)\right\},
 P_{\mathcal{I}}^{-h}(w_{\mathcal{I}})\right).
\end{equation}
Since both $i^+$ and $j^-$ belong to $\cal J$, the momentum $P_{\cal
I}$ does not depend on $z$. Likewise from the definition of $w_{\cal
I}$ in (\ref{pole}) one can see that it does not depend on $z$.
Therefore, the $z$ dependence is confined to the second amplitude in
(\ref{fiki}) which we can write more explicitly as
\begin{equation}
\label{amlu}
M_{\mathcal{J}}\left(\left\{r_{\mathcal{J}'}^{-}\right\},
\left\{k_{\mathcal{J}'}^{+}(w_{\mathcal{I}})\right\},
\{\lambda^{(i)}(w_{\cal I},z),\tilde\lambda^{(i)}\},
 \{\lambda^{(j)},\tilde\lambda^{(j)}(w_{\cal I},z)\},
 P_{\mathcal{I}}^{-h}(w_{\mathcal{I}})\right)
\end{equation}
where the set ${\cal J}' = {\cal J}\setminus \{i,j\}$. It is
straightforward to show that
\begin{equation}
\lambda^{(i)}(w_{\cal I},z) = \lambda^{(i)}(w_{\cal I}) +
z\lambda^{(j)}, \qquad \tilde\lambda^{(j)}(w_{\cal I},z) =
\tilde\lambda^{(j)}(w_{\cal I}) - z\tilde\lambda^{(i)}.
\end{equation}

The fact that $\lambda^{(i)}(w_{\cal I})$ and
$\tilde\lambda^{(j)}(w_{\cal I})$ get deformed exactly in the same
way as $\lambda^{(i)}$ and $\tilde\lambda^{(j)}$ do is what allows
us to use induction for these terms. Note that the amplitude
(\ref{amlu}) is therefore a physical amplitude with a BCFW
deformation. The number of gravitons is less than $n$ and by our
induction hypothesis it vanishes as $z$ goes to infinity.

To complete the induction argument it suffices to note that the
auxiliary recursion relations we are using can reduce any amplitude
to products of three graviton amplitudes. Finally, recall that the
Feynman diagram argument at the beginning of this section showed
that amplitudes with less than five gravitons vanish at infinity
under the BCFW deformation.

Consider now a term of the second kind,
\begin{equation}\label{sasi}
\begin{split}
\sum_{h=\pm} &
M_{\mathcal{I}}\left(\left\{r_{\mathcal{I}}^{-}\right\},
 \left\{k_{\mathcal{I}}^{+}(w_{\mathcal{I}}(z),z)\right\},
 -P_{\mathcal{I}}^{h}(w_{\mathcal{I}}(z),z)\right)\frac{1}{P^{2}_{\mathcal{I}}(z)}
 \\ &
M_{\mathcal{J}}\left(\left\{r_{\mathcal{J}}^{-}(w_{\cal
I}(z),z)\right\},
 \left\{k_{\mathcal{J}}^{+}(w_{\mathcal{I}}(z))\right\},
 P_{\mathcal{I}}^{-h}(w_{\mathcal{I}}(z),z)\right).
\end{split}
\end{equation}
Recall that for these terms $i^+\in {\cal I}$ while $j^-\in {\cal
J}$. The $z$ dependence we have displayed in (\ref{sasi}) looks
complicated at first since
\begin{equation}
w_{\cal I}(z)\:=\:\frac{P_{\mathcal{I}}(z)^{2}}{
       \sum_{k\in\mathcal{I}^+}\left<j|P_{\mathcal{I}}(z)|k\right]
       }
\end{equation}
appears to be a rational function of $z$ since $P_{\cal I}(z)_{a\dot
a} = P_{{\cal I}\: a\dot a}
+z\lambda^{(j)}_a\tilde\lambda^{(i)}_{\dot a}$. Note, however, that
$\tilde\lambda^{(k)}$'s with $k\in {\cal I}^+$ do not depend on $z$
and that the $z$ dependence
$z\lambda^{(j)}_a\tilde\lambda^{(i)}_{\dot a}$ in $P_{\cal I}(z)$
drops out of the denominator thanks to the contraction with $\langle
j|$.

Then we find that $w_{\cal I}(z)$ is simply a linear function of
$z$:
\begin{equation}
\label{amus}
w_{\cal I}(z)\:=\: w_{\cal I} - z \left( \frac{\left<
j|P_{\mathcal{I}}|i\right]}{
       \sum_{k\in\mathcal{I}^+}\left<j|P_{\mathcal{I}}|k\right]}
       \right)
\end{equation}
where $w_{\cal I}$ is just the undeformed one, i.e., $w_{\cal
I}(0)$.

The final step before we proceed to study the behavior for $z\to
\infty$ using Feynman diagrams is to determine the properties of the
internal graviton that enters with opposite helicities in the
amplitudes of (\ref{sasi}). The momentum of the internal graviton is
given by
\begin{equation}
P_{\cal I}(w_{\cal I}(z),z) = \sum_{k\in {\cal I}^-}p_k +p_i(w_{\cal
I}(z),z) + \sum_{s\in {\cal I}^+,\; s\neq i} p_s(w_{\cal I}(z)).
\end{equation}

The important observation is that the $z$-dependence can be fully
separated as follows
\begin{equation}
\label{veci} P_{\cal I}(w_{\cal I}(z),z) = P_{\cal I}(w_{\cal I}) +
z\lambda^{(j)} \left( - \left( \frac{\left<
j|P_{\mathcal{I}}|i\right]}{
       \sum_{k\in\mathcal{I}^+}\left<j|P_{\mathcal{I}}|k\right]}
       \right)\sum_{s\in {\cal I}^+}\tilde\lambda^{(s)} + \tilde\lambda^{(i)}
\right)
\end{equation}
where $P_{\cal I}(w_{\cal I})$ is the $z$-undeformed one, i.e.,
$P_{\cal I}(w_{\cal I}(0),0)$.

Note that we have written $P_{\cal I}(w_{\cal I}(z),z)$, which is a
null vector, as the sum of two null vectors. For real momenta, this
would imply that all three vectors are proportional. However, in
this case all three vectors are complex and all that is required is
that either all $\lambda$'s or all $\tilde\lambda$'s be
proportional. We claim that in this particular case all
$\tilde\lambda$'s are proportional. To see this note that if we
write $P_{\cal I}(w_{\cal I})_{a\dot a} =
\lambda^{(P)}_a\tilde\lambda^{(P)}_{\dot a}$, then
$\tilde\lambda^{(P)}_{\dot a}$ is proportional to $\zeta_{\dot a} =
\eta^{a}P_{\cal I}(w_{\cal I})_{a\dot a}$ for some arbitrary spinor
$\eta^{a}$.

We claim that the $\tilde\lambda$ spinor of the vector multiplying
$z$ in (\ref{veci}) is also proportional to $\zeta^{\dot a}$ if
$\eta_a = \lambda_a^{(j)}$. In this case, $\zeta_{\dot a} =
\lambda^{(j)\; a}P_{\cal I}(w_{\cal I})_{a\dot a}= \lambda^{(j)\;
a}P_{{\cal I}\; a\dot a}$. To prove our claim consider the inner
product of the two spinors
\begin{equation}
\label{cane} \left( -\left( \frac{\left<
j|P_{\mathcal{I}}|i\right]}{
       \sum_{k\in\mathcal{I}^+}\left<j|P_{\mathcal{I}}|k\right]}
       \right)\sum_{s\in {\cal I}^+}\tilde\lambda^{(s)}_{\dot a} +
       \tilde\lambda^{(i)}_{\dot a}
\right)\zeta^{\dot a} =\left( \left( \frac{\left<
j|P_{\mathcal{I}}|i\right]}{
       \sum_{k\in\mathcal{I}^+}\left<j|P_{\mathcal{I}}|k\right]}
       \right)\sum_{s\in {\cal I}^+}\langle j | P_{\cal I} |s ]  -
       \langle j | P_{\cal I} |i ]
\right).
\end{equation}

The right hand side of (\ref{cane}) vanishes trivially showing that
the two spinors are proportional.

Therefore, it follows that we can write $P_{\cal I}(w_{\cal
I}(z),z)_{a\dot a} = \lambda_a(z)\tilde\lambda^{P}_{\dot a}$ where
$\lambda_a(z) = \lambda_a^{(P)} + z \beta \lambda^{(j)}_a$ for some
$\beta$ which is $z$ independent. Note that if $z=0$ we recover
$P_{\cal I}(w_{\cal I})_{a\dot a} =
\lambda^{(P)}_a\tilde\lambda^{(P)}_{\dot a}$.

Let us turn to the analysis of the amplitudes in (\ref{sasi}) to
show that their product vanishes as $z$ is taken to infinity. In
other words, we will see that $M_{\cal I}$ and $M_{\cal J}$ may not
vanish simultaneously but their product together with the propagator
always does.

Consider the first amplitude
$M_{\mathcal{I}}\left(\left\{r_{\mathcal{I}}^{-}\right\},
 \left\{k_{\mathcal{I}}^{+}(w_{\mathcal{I}}(z),z)\right\},
 -P_{\mathcal{I}}^{h}(w_{\mathcal{I}}(z),z)\right)$. Let
 the number of particles in the sets $\{ r_{\cal I}^-\}$ and $\{ k_{\cal
 I}^+\}$ be $m_{\cal I}$ and $p_{\cal I}$ respectively\footnote{
Note that if $h=+$ this is a physical amplitude where only the
$\lambda$'s of positive helicity gravitons have been deformed. It is
interesting to note that this deformation is basically the one
introduced by Risager in \cite{Risager:2005vk} and later in
\cite{Bjerrum-Bohr:2005jr} to construct an MHV diagram expansion for
gravity amplitudes.}.

The Feynman diagram analysis is very similar to that performed at
the beginning of section III. The leading Feynman diagram is again
one with only cubic vertices that posses a quadratic dependence on
momenta. The number of cubic vertices is the total number of
particles\footnote{The total number of gravitons in $M_{\cal I}$ is
$m_{\cal I}+p_{\cal I}+1$ since $-P^h_{\cal I}(w_{\cal I}(z),z)$
should also be included.} minus two, i.e, $m_{\cal I}+p_{\cal I}-1$.
Therefore the contribution from vertices gives at most a factor of
$z^{2(m_{\cal I}+p_{\cal I}-1)}$. There are $p_{\cal I}+1$
polarization vectors that depend on $z$, giving a total contribution
of $1/z^{2(p_{\cal I}+h)}$. Here we have used that since $z$ enters
in $-P_{\mathcal{I}}^{h}(w_{\mathcal{I}}(z),z)$ only through
$\lambda(z)$, its polarization tensor gives a contribution of
$1/z^{2h}$. Finally, we need to count the number of propagators that
depend on $z$. It turns out that there are exactly $m_{\cal
I}+p_{\cal I}-2$ of them giving a contribution of $1/z^{m_{\cal
I}+p_{\cal I}-2}$. This last statement is not obvious since there
could be accidental cancelations of the $z$ dependence. Let us
continue with the argument here and we will prove that there is no
accidental cancelations within the propagators in the next
subsection\footnote{More precisely, what we prove in the next
subsection is that trivial cancelations in which neither propagators
nor vertices depend on $z$ are the only ones that can occur.}.
Collecting all factors we get
\begin{equation}
\label{uno} M_{\mathcal{I}}\left(\left\{r_{\mathcal{I}}^{-}\right\},
 \left\{k_{\mathcal{I}}^{+}(w_{\mathcal{I}}(z),z)\right\},
 -P_{\mathcal{I}}^{h}(w_{\mathcal{I}}(z),z)\right) \sim \frac{1}{z^{p_{\cal I}-m_{\cal
 I}+2h}}.
\end{equation}

The propagator $1/P^{2}_{\mathcal{I}}(z)$ in (\ref{sasi}) goes as
$1/z$.

The reader might have noticed that in this argument special care is
required when ${\cal I}^+ =\{i\}$. We postpone the study of this
case to the end of the section. Until then we simply assume that
$i\in {\cal I}^+$ but ${\cal I}^+ \neq \{i\}$.

Consider now the second amplitude in (\ref{sasi}),
\begin{equation}
M_{\mathcal{J}}\left(\left\{r_{\mathcal{J}}^{-}(w_{\cal
I}(z),z)\right\},
 \left\{k_{\mathcal{J}}^{+}(w_{\mathcal{I}}(z))\right\},
 P_{\mathcal{I}}^{-h}(w_{\mathcal{I}}(z),z)\right).
\end{equation}
Let the number of gravitons in $\{ r_{\cal J}^-\}$ and $\{ k_{\cal
 J}^+\}$ be $m_{\cal J}$ and $p_{\cal J}$ respectively.

The cubic vertices give again a factor of $z^{2(p_{\cal J}+m_{\cal
J}-1)}$. The polarization tensors give a factor of $1/z^{2(p_{\cal
J}-h+1)}$. Here we have taken into account the contribution from the
$z$ dependent negative helicity graviton, i.e, the $j^{\rm th}$
graviton, and from the internal graviton,
$P_{\mathcal{I}}^{-h}(w_{\mathcal{I}}(z),z)$. Finally, the
propagators contribute again a factor of $1/z^{p_{\cal J}+m_{\cal
J}-2}$. Collecting all factors we get
\begin{equation}
\label{tres} M_{\mathcal{J}}\left(\left\{r_{\mathcal{J}}^{-}(w_{\cal
I}(z),z)\right\},
 \left\{k_{\mathcal{J}}^{+}(w_{\mathcal{I}}(z))\right\},
 P_{\mathcal{I}}^{-h}(w_{\mathcal{I}}(z),z)\right) \sim \frac{1}{z^{p_{\cal J}-m_{\cal
 J}-2(h-1)}}.
\end{equation}
Combining all contributions from (\ref{uno}), the propagator and
(\ref{tres}), the leading $z$ behavior of (\ref{sasi}) is
$1/z^{p-m+3}$.

This shows that all the amplitudes with $p\ge m$ vanish at infinity.

As stated at the beginning of this subsection, a similar discussion
holds for the case of amplitudes with $m\ge p$: by repeating the
same counting starting from relation (\ref{RRminus}), the behavior
at infinity of terms of the second kind turns out to be
$1/z^{m-p+3}$. Terms of the first kind can again be treated by
induction.

It is important to mention that the way amplitudes vanish at
infinity is generically only as $1/z^2$. This is because terms of
the first kind which are treated by induction vanish as
three-graviton amplitudes do, i.e, as $1/z^2$.

This completes our proof of the vanishing of $M_n(z)$ as $z$ goes to
infinity up to the claim made about the number of propagators that
contribute a $1/z$ factor and the exceptional case when ${\cal I}^+
= \{ i \}$. We now turn to these crucial steps of our proof.

\subsection{Analysis Of The Contribution From Propagators}

One thing left to prove is that in the leading Feynman diagrams
contributing to the first amplitude, $M_{\mathcal{I}}$, there are
exactly $m_{\cal I}+p_{\cal I}-2$ propagators giving a $1/z$
contribution at infinity while in the second amplitude,
$M_{\mathcal{J}}$, there are exactly $m_{\cal J}+p_{\cal J}-2$ of
them.

\subsubsection{Propagators In Leading Feynman Diagrams Of $M_{\cal I}$}

Let us start with $M_{\cal I}$. The argument here uses similar
elements to the ones given in the appendix where we provided a proof
of the auxiliary recursion relations.

Consider a given Feynman diagram. A propagator naturally divides the
diagram into two subdiagrams. Let use denote them by ${\cal L}$ and
${\cal R}$. Without loss of generality, we can always take the
graviton with momentum $-P^h_{\cal I}(w_{\cal I}(z),z)$ to be in
${\cal R}$. In the set of positive helicity gravitons, $\{ k_{\cal
I}^+(w_{\cal I}(z),z)\}$, there is one that is special; the $i^{\rm
th}$ graviton. We consider two cases, the first is when $i\in {\cal
L}^+$ and the second when $i\in {\cal R}^+$.

\smallskip

{\it Case A: $i\in {\cal L}^+$}

\smallskip

Let $i\in {\cal L}^+$, then the propagator under consideration has
the form
\begin{equation}
P_{\cal L}(w_{\cal I}(z),z) = P_{\cal L}(w_{\cal I}(0)) +
z\lambda^{(j)}\left( -\frac{\langle j|P_{\cal I}|i]}{\sum_{k\in
{\cal I}^+}\langle j|P_{\cal I}|k]}\sum_{s\in {\cal
L}^+}\tilde\lambda^{(s)} + \tilde\lambda^{(i)}\right).
\end{equation}
We are interested in asking when
\begin{equation}
\label{propa} P_{\cal L}(w_{\cal I}(z),z)^2 =P_{\cal L}(w_{\cal
I}(0))^2 + z\left( \frac{\langle j|P_{\cal I}|i]\sum_{k\in {\cal
L}^+}\langle j|P_{\cal L}|k]}{\sum_{k\in {\cal I}^+}\langle
j|P_{\cal I}|k]} -\langle j|P_{\cal L}|i] \right)
\end{equation}
can be $z$ independent. Therefore we have to analyze under which
conditions the factor multiplying $z$ can be zero for a generic
choice of momenta and polarization tensors of the physical gravitons
subject only to the overall momentum conservation constrain.

Let us write the factor of interest as follows
\begin{equation}
\label{coso} \langle j|P_{\cal I}|i]\sum_{k\in {\cal L}^+}\langle
j|P_{\cal L}|k] -\langle j|P_{\cal L}|i]\sum_{k\in {\cal
I}^+}\langle j|P_{\cal I}|k] = \lambda^{(j)\; a}\lambda^{(j)\;
b}P_{{\cal L}\; a\dot a}P_{{\cal I}\; b\dot b}T^{\dot a\dot b}
\end{equation}
with
\begin{equation}
T^{\dot a\dot b} = \tilde\lambda^{(i)\; \dot a}\sum_{k\in {\cal
L}^+}\tilde\lambda^{(k)\; \dot b} - \tilde\lambda^{(i)\; \dot
b}\sum_{k\in {\cal I}^+}\tilde\lambda^{(k)\; \dot a}.
\end{equation}

Here we have to consider two different cases\footnote{There are
actually three cases. The third is when ${\cal I}^+ = {\cal L}^+ =\{
i \}$ but this is part of the special case that is considered at
then end of the section.}:

\begin{itemize}

\item ${\cal I}^+\setminus {\cal L}^+ \neq {\emptyset}$.

\item ${\cal I}^+ = {\cal L}^+$ and ${\cal L}^+ \neq \{ i \}$.

\end{itemize}

Let us start by assuming that ${\cal I}^+\setminus {\cal L}^+$ is
non-empty and that, say, $s\in {\cal I}^+\setminus {\cal L}^+$. The
space of kinematical invariants we consider is determined by the
momentum and polarization tensors of each of the original gravitons.
Consider both objects for the $s^{\rm th}$ graviton
\begin{equation}
\label{muka} \epsilon^{+\; (s)}_{a\dot a, b \dot b} =
\frac{\mu_a\tilde\lambda^{(s)}_{\dot
a}\mu_b\tilde\lambda^{(s)}_{\dot b}}{\langle \mu,
\lambda^{(s)}\rangle^2}, \qquad p^{(s)}_{a\dot a} =
\lambda_a^{(s)}\tilde\lambda_{\dot a}^{(s)}.
\end{equation}
It is clear that if we take $\{ \lambda^{(s)}_a,
\tilde\lambda^{(s)}_{\dot a}\}$ to $\{ t^{-1}\lambda^{(s)}_a,
t\tilde\lambda^{(s)}_{\dot a}\}$ with $t$ a fourth root of unity,
i.e, $t^4=1$ then (\ref{muka}) is invariant. Therefore, any quantity
that vanishes for $t=1$ must also vanish for all four values of $t$.
In particular, it must be the case that (\ref{coso}) must vanish for
all four values of $t$. Since momentum is not affected only the
tensor $T^{\dot a\dot b}$ changes. Taking the difference between two
values of $t$, say $t=1$ and $t=i$, we find that $T^{\dot a\dot
b}|_{t=1} - T^{\dot a\dot b}|_{t=i} \sim \tilde\lambda^{(i)\; \dot
b}\tilde\lambda^{(s)\; \dot a}$. Therefore, the vanishing of
(\ref{coso}) implies that of
\begin{equation}
\langle j|P_{\cal L}|i]\langle j|P_{\cal I}|s] = 0.
\end{equation}
This condition is then equivalent to
\begin{equation}
{\rm tr}\left( \slash \!\!\! p_j \; \slash \!\!\!\! P_{\cal L}\;
\slash \!\!\! p_i \; \slash \!\!\!\! P_{\cal L}\right) = 0 \qquad
{\rm or} \qquad {\rm tr}\left( \slash \!\!\! p_j\; \slash \!\!\!\!
P_{\cal I}\; \slash \!\!\! p_s\; \slash \!\!\!\! P_{\cal I}\right) =
0
\end{equation}
but these are constraints on the kinematical space which are not
satisfied at generic points.

The second case we have to consider is when ${\cal I}^+ = {\cal
L}^+$ and ${\cal L}^+ \neq \{ i \}$. Let us introduce the notation
$\tilde\mu_{\dot a} =\sum_{k\in {\cal I}^+}\tilde\lambda_{\dot
a}^{(k)}$. Therefore the condition we want to exclude is
\begin{equation}
\langle j|P_{\cal I}|i]\langle j|P_{\cal L}|\tilde\mu] -\langle
j|P_{\cal L}|i]\langle j|P_{\cal I}|\tilde\mu] = 0.
\end{equation}
Using Schouten's identity we can write this as
\begin{equation}
\langle j | P_{\cal I} P_{\cal L} |j\rangle [i,\tilde\mu] = 0.
\end{equation}
The vanishing of either factor\footnote{See section 1.A for the
explanation of the notation in the first factor.} implies a
constraint for the space of kinematical invariants. In the case of
the second factor this can easily be seen by choosing $s\in {\cal
I}^+$ and $s\neq i$, then using the scaling by $t$ with $t^4=1$ to
conclude that $(p_s+p_i)^2=0$.

This completes the proof that the $z$ dependence cannot drop out of
any propagator and therefore all $m_{\cal I}+p_{\cal I}-2$ of them
give a $1/z$ factor in $M_{\cal I}$ if $i\in {\cal L}$.

\smallskip

{\it Case B: $i\in {\cal R}^+$}

\smallskip

The analysis when $i\in {\cal R}$ is completely analogous except for
the fact that there is one case that was not possible before. As we
will show, this will correspond to diagrams which give a non-leading
contribution.

Consider the analog of (\ref{propa})
\begin{equation}
P_{\cal L}(w_{\cal I}(z),z)^2 =P_{\cal L}(w_{\cal I}(0))^2 +
z\langle j|P_{\cal I}|i]\left( \frac{\sum_{k\in {\cal L}^+}\langle
j|P_{\cal L}|k]}{\sum_{k\in {\cal I}^+}\langle j|P_{\cal I}|k]}
\right).
\end{equation}
The new case is when ${\cal L}^+ = \emptyset$, then the $z$
dependence drops out. Of course, this is not a problem because if
the set ${\cal L}^+$ is empty it means that nothing on the
subdiagram ${\cal L}$ depends on $z$, including the cubic vertices.
Therefore, neither propagators nor cubic vertices contribute. One
can then concentrate on the subdiagram ${\cal R}$, but this
subdiagram has less particles than the total diagram and the same
number of $z$-dependent polarization tensors. Therefore these
diagrams go to zero even faster than diagrams where ${\cal L}^+$ is
not empty.

\subsubsection{Propagators In Leading Feynman Diagrams Of $M_{\cal J}$}

Let us now study the leading Feynman diagrams contributing to
$M_{\cal J}$. Again, the propagator divides the diagram in two
subdiagrams that we denote $\cal L$ and $\cal R$. Without loss of
generality, we can always take the graviton with momentum
$P^{-h}_{\cal I}(w_{\cal I}(z),z)$ to be in ${\cal R}$. As in the
previous discussion we have a special graviton, i.e, the $j^{\rm
th}$ graviton. Therefore we have to consider two cases, $j\in {\cal
L}$ and $j\in {\cal R}$.

\smallskip

{\it Case A: $j\in {\cal L}^-$}

\smallskip

Let us first consider the case $j\in {\cal L}$. The $z$ dependence
of $\tilde\lambda^{(j)}(w_{\cal I}(z),z)$ is the most complicated of
all. This is why we write it explicitly
\begin{equation}
\tilde\lambda^{(j)}(w_{\cal I}(z),z)_{\dot a} =
\tilde\lambda^{(j)}(w_{\cal I})_{\dot a} + z\left(
-\tilde\lambda^{(i)}_{\dot a} + \frac{\langle j|P_{\cal
I}|i]}{\sum_{k\in {\cal I}^+}\langle j|P_{\cal I}|k]}\sum_{s\in \{
k^+\} }\tilde\lambda^{(s)}_{\dot a} \right).
\end{equation}
Using this and the fact that the set of labels of all positive
helicity gravitons $\{k^+\}$ must be equal to ${\cal I}^+\cup {\cal
J}^+$, we find that the propagator of interest has a momentum
dependence of the form
\begin{equation}
P_{\cal L}(w_{\cal I}(z),z)^2 =P_{\cal L}(w_{\cal I}(0))^2 + z\left(
\langle j|P_{\cal L}|i]- \langle j|P_{\cal I}|i]\frac{\sum_{k\in
{\cal I}^+\cup ({\cal J}^+\setminus {\cal L}^+)}\langle j|P_{\cal
L}|k]}{\sum_{k\in {\cal I}^+}\langle j|P_{\cal I}|k]} \right).
\end{equation}
We are then interested in asking when this expression can be $z$
independent.

The analysis is similar to the one given for $M_{\cal I}$ so we will
be brief. The factor of interest is now
\begin{equation}
\label{moji} \langle j|P_{\cal L}|i]\sum_{k\in {\cal I}^+}\langle
j|P_{\cal I}|k] -\langle j|P_{\cal I}|i]\sum_{k\in {\cal I}^+\cup
({\cal J}^+\setminus {\cal L}^+)}\langle j|P_{\cal L}|k].
\end{equation}

We have to consider two cases:

\begin{itemize}

\item ${\cal J}^+\setminus {\cal L}^+ \neq {\emptyset}$.

\item ${\cal J}^+ = {\cal L}^+$ and ${\cal I}^+\neq \{i\}$.

\end{itemize}

In the first case we can assume that, say, the $s^{\rm th}$ graviton
is in ${\cal J}^+\setminus {\cal L}^+$. Then by using the argument
that any statement about $\{ \lambda^{(s)},\tilde\lambda^{(s)}\}$
must also be true for $\{
t^{-1}\lambda^{(s)},t\tilde\lambda^{(s)}\}$ with $t^4=1$ one can
show that the vanishing of (\ref{moji}) implies a nontrivial
constraint on kinematical invariants that is not generically
satisfied.

The second case is also similar to one we considered in the analysis
of $M_{\cal I}$. Here we have that ${\cal I}^+\cup ({\cal
J}^+\setminus {\cal L}^+)={\cal I}^+\cup \emptyset = {\cal I}^+$.
Therefore (\ref{moji}) becomes
\begin{equation}
\langle j|P_{\cal L}|i]\langle j|P_{\cal I}|\tilde\mu] -\langle
j|P_{\cal I}|i]\langle j|P_{\cal L}|\tilde\mu]
\end{equation}
where $\tilde\mu_{\dot a} = \sum_{s\in {\cal I}^+\setminus \{ i
\}}\tilde\lambda^{(s)}$. Since by assumption ${\cal I}^+\setminus \{
i \}\neq \emptyset$ we can use Schouten's identity to derive
non-trivial constraints on the kinematical invariants which are not
satisfied for generic momenta.

Recall that the case when ${\cal I}^+ = \{ i \}$ is special and will
be treated separately.

\smallskip

{\it Case B: $j\in {\cal R}^-$}

\smallskip

In this case, the propagator of interest can be written as
\begin{equation}
P_{\cal L}(w_{\cal I}(z),z)^2 =P_{\cal L}(w_{\cal I}(0))^2 +
z\langle j|P_{\cal I}|i]\left( \frac{\sum_{k\in {\cal L}^+}\langle
j|P_{\cal L}|k]}{\sum_{k\in {\cal I}^+}\langle j|P_{\cal I}|k]}
\right).
\end{equation}
This is again similar to the corresponding case in $M_{\cal I}$. The
only new case compared to when $j\in {\cal L}^-$ is when ${\cal
L}^+$ is empty. Then nothing in ${\cal L}$ depends on $z$ and we can
consider a Feynman diagram that has less minus helicity gravitons
than the original one and therefore it goes faster to zero at
infinity than the leading diagrams obtained when ${\cal L}^+\neq
\emptyset$.

This conclude our discussion about the contribution of the
propagators.

\subsection{Analysis Of The Special Case ${\cal I}^+ = \{ i\}$}

Let us now consider the final case. This is when ${\cal I}^+ = {\cal
L}^+ = \{ i \}$. This case is quite interesting since several
unexpected cancelations take place. Consider $w_{\cal I}(z)$ given
in (\ref{amus}). In this case, it is easy to check that $w_{\cal
I}(z) = w_{\cal I}(0) - z$. A consequence of this is that
$\lambda^{(i)}(w_{\cal I}(z),z) = \lambda^{(i)}(z)+w_{\cal
I}(z)\lambda^{(j)}$ becomes $z$-independent. To see this recall that
$\lambda^{(i)}(z) = \lambda^{(i)} + z\lambda^{(j)}$. Therefore
$\lambda^{(i)}(w_{\cal I}(z),z) =\lambda^{(i)}(w_{\cal I})$. This
also implies that $P_{\cal I}^h(w_{\cal I}(z),z)$ is $z$
independent. Therefore, the full amplitude $M_{\cal I}$ is $z$
independent.

Recall that we are interested in the behavior of
\begin{equation}
\sum_{h=\pm}M_{\cal I}^h\frac{1}{P^2_{\cal I}(z)}M_{\cal J}^{-h}(z).
\end{equation}
The propagator $1/P^2_{\cal I}(z)$ contributes a factor of $1/z$.

Now we have to look at
$$M_{\cal J}(z) = M_{\mathcal{J}}\left(\left\{r_{\mathcal{J}}^{-}(w_{\cal
I}(z),z)\right\},
 \left\{k_{\mathcal{J}}^{+}(w_{\mathcal{I}}(z))\right\},
 P_{\mathcal{I}}^{-h}(w_{\mathcal{I}}(z),z)\right).$$
Let us study the $z$ dependence of each graviton carefully. We have
that the $j^{\rm th}$ graviton (which has negative helicity) and all
positive helicity gravitons in ${\cal J}^+
=\left\{k_{\mathcal{J}}^{+}(w_{\mathcal{I}}(z))\right\}$ behave as
\begin{equation}
\label{wowi} \tilde\lambda^{(j)}(w_{\cal I}(z),z) =
\tilde\lambda^{(j)}(w_{\cal I}) + z \sum_{s\in {\cal
J}^+}\tilde\lambda^{(s)}, \qquad \lambda^{(s)}(w_{\cal I}(z)) =
\lambda^{(s)}(w_{\cal I}) - z\lambda^{(j)} \qquad \forall \; s\in
{\cal J}^+.
\end{equation}
Close inspection of (\ref{wowi}) shows a striking fact. This
deformation is exactly the same as the one that led to the auxiliary
recursion relations in the first place, i.e, the deformation given
in (\ref{csd}) but using $z$ instead of $w$ as deformation parameter
and $\tilde\lambda^{(j)}(w_{\cal I})$ and $\lambda^{(s)}(w_{\cal
I})$ as undeformed spinors. Finally, recall that
$P_{\mathcal{I}}(w_{\mathcal{I}}(z),z)$, which also appears in
$M_{\cal J}$, was shown to be $z$ independent.

Now, if $h=+$ we have $P_{\mathcal{I}}^{-}(w_{\mathcal{I}})$ and
therefore, $M_{\cal J}(z)$ is nothing but a physical amplitude under
the maximal deformation (\ref{csd}). In the appendix, we showed that
amplitudes vanish as the deformation parameter, which in this case
is $z$, is taken to infinity if the number of pluses is greater than
or equal to the number of minuses minus two. To see that this
condition is satisfied in $M_{\cal J}$ note that since ${\cal I}^+ =
\{ i\}$ we have that the total number of positive helicity gravitons
in $M_{\cal J}$ is $p-1$ while that of negative helicity gravitons
is $m - m_{\cal I}+1$. Since the number of external negative
helicity gravitons in $M_{\cal I}$ must be at least one, i.e,
$m_{\cal I}\geq 1$ and recalling that we are studying the case when
$p\geq m$, we get the desired result.

The next case to consider is when $h=-$. Since
$P_{\mathcal{I}}^{+}(w_{\mathcal{I}})$ is $z$ independent, the
deformation (\ref{wowi}) of $M_{\cal J}$ is no longer maximal.
However, it is possible to show that these terms are identically
zero. This is obvious when the on-shell physical amplitude
$M_{\mathcal{I}}$, which has only one positive helicity graviton,
has more than two negative helicity gravitons.

Consider now the case when $M_{\mathcal{I}}$ has precisely two
negative helicity gravitons. A three-graviton on-shell amplitude
need not vanish if momenta are complex therefore this is a
potentially dangerous case. Three-graviton amplitudes are given as
the square of the gauge theory ones. Therefore we have
\begin{equation}\label{MHVMI}
M_{\mathcal{I}}(i^{+}(w_{\mathcal{I}}),s^{-},-P_{\mathcal{I}}^{-}
                (w_{\mathcal{I}}))\:=\:
 \left(
  \frac{\langle \lambda^{(s)}, \lambda^{(P)}\rangle^{3}}{
        \langle \lambda^{(P)}, \lambda^{(i)}(w_{\mathcal{I}})\rangle
        \langle \lambda^{(i)}(w_{\mathcal{I}}), \lambda^{(s)}\rangle}
 \right)^2
\end{equation}
where as in section III.C we have defined $P_{\cal I}(w_{\cal
I})_{a\dot a} =\lambda^{(P)}_a\tilde\lambda^{(P)}_{\dot a}$.

Since this is a physical amplitude, momentum is conserved which
means
\begin{equation}\label{PIw}
 \lambda^{(i)}(w_{\mathcal{I}})_a\tilde\lambda^{(i)}_{\dot a} +
 \lambda^{(s)}_a\tilde\lambda^{(s)}_{\dot
 a}\:=\:\lambda^{(P)}_a
  \tilde\lambda^{(P)}_{\dot a}.
\end{equation}
For real momenta, this equation implies that all $\lambda's$ and all
$\tilde\lambda's$ are proportional. Therefore three-graviton
amplitudes must vanish. For complex momenta, this need not be the
case and one can have all $\tilde\lambda's$ be proportional with the
$\lambda$'s unconstrained. In such a case (\ref{MHVMI}) would not
vanish.

We claim that, luckily in our case of interest, all $\lambda's$ are
proportional and (\ref{MHVMI}) vanishes. To see this note that
$w_{\cal I} = -\langle i, s \rangle /\langle j, s \rangle $ and
$\lambda^{(i)}(w_{\cal I})_{a} = \lambda^{(i)}_a + w_{\cal
I}\lambda^{(j)}_a$, therefore $\langle \lambda^{(i)}(w_{\cal I}),
\lambda^{(s)} \rangle =0$. Contracting (\ref{PIw}) with
$\lambda^{(s)\; a}$ we find $\langle \lambda^{(P)}, \lambda^{(s)}
\rangle \tilde\lambda^{(P)}_{\dot a} = 0$. Therefore we must have
$\langle \lambda^{(P)}, \lambda^{(s)} \rangle = 0$ which completes
the proof of our claim.

From (\ref{MHVMI}), this condition implies that $M_{\mathcal{I}}$ is
identically zero. Thus, we can conclude that the cases of $M_{\mathcal{J}}$
with a non-maximal deformation are not there.

This is the end of our proof. We now turn to some extensions and
applications of the BCFW recursion relations that can be obtained by
using Ward identities.

\section{Ward Identities}

Our proof of the BCFW recursion relations was based on deforming two
gravitons of opposite helicities, $i^+$ and $j^-$, in the following
way:
\begin{equation}
\label{bcfwz} \lambda^{(i)}(z) = \lambda^{(i)} + z\lambda^{(j)},
\qquad \tilde\lambda^{(j)}(z) = \tilde\lambda^{(j)} -
z\tilde\lambda^{(i)}.
\end{equation}
However, it is known that in gauge theory, deformed amplitudes also
vanish at infinity if the helicities $(h_{i},h_{j})$ of the deformed
gluons are $(-,-)$ or $(+,+)$ \cite{Britto:2005fq}. It would be
interesting to prove a similar statement for General Relativity.
Here we show that this is indeed very straightforward in the case of
MHV scattering amplitudes if one uses Ward identities.

The Ward identity of relevance for our discussion can be found for
example in \cite{Bern:2002kj} and it is given by
\begin{equation}
\label{wardi} \frac{M^{\mbox{\tiny MHV}}_{l,m}}{\langle
\lambda^{(l)},\lambda^{(m)}\rangle^{8}} = \frac{M^{\mbox{\tiny
MHV}}_{s,q}}{\langle \lambda^{(s)},\lambda^{(q)}\rangle^{8}},
\end{equation}
where the notation $M^{\mbox{\tiny MHV}}_{a,b}$ indicates that the
gravitons $a$ and $b$ in this amplitude are the ones with negative
helicity.

Consider first the $(+,+)$ case. We use the Ward identity
(\ref{wardi}) to relate it to the usual $(+,-)$ case. For clarity
purposes, we explicitly exhibit the dependence of the amplitudes on
only four gravitons: $\{ l,m,i,j\}$. The dependence on the rest of
the gravitons (all of which have positive helicity) will be
implicit. Then we have
\begin{equation}\label{WardMHV1}
M_{n}^{\mbox{\tiny MHV}}(i^{+} (z),j^{+}(z),l^{-},
 m^{-})\:=\:
 \left(\frac{\langle \lambda^{(l)},\lambda^{(m)}\rangle}{\langle
 \lambda^{(j)},\lambda^{(l)}\rangle}\right)^8
M_{n}^{\mbox{\tiny MHV}}(i^{+}(z),j^{-}(z),l^{-},
 m^{+}).
\end{equation}
The MHV amplitude on the right hand-side is deformed as in
(\ref{bcfwz}), thus it vanishes at infinity by our proof. Since both
inner products expressed explicitly in (\ref{WardMHV1}) do not
depend on $z$, the amplitude on the left hand side of
(\ref{WardMHV1}), where $(h_{i},h_{j})\,=\,(+,+)$, will vanish as
$z$ goes to infinity.

Consider now the $(-,-)$ case. Using again the Ward identity
(\ref{wardi}) we have
\begin{equation}\label{WardMHV2}
M_{n}^{\mbox{\tiny MHV}}(i^{-} (z),j^{-}(z),l^{+},
 m^{+})\:=\:
 \left( \frac{\langle \lambda^{(i)}(z),\lambda^{(j)}\rangle}
 {\langle \lambda^{(j)},\lambda^{(l)}\rangle}\right)^8
M_{n}^{\mbox{\tiny MHV}}(i^{+}(z),j^{-}(z),l^{-},
 m^{+})
\end{equation}
Note that $\langle \lambda^{(i)}(z),\lambda^{(j)}\rangle$ does not
depend on $z$ since $\lambda^{(i)}(z) = \lambda^{(i)} +
z\lambda^{(j)}$. Therefore, the amplitude still vanishes in this
case.

In \cite{Bedford:2005yy}, a very nice compact formula was
conjectured for MHV amplitudes of gravitons by assuming the validity
of BCFW recursion relations obtained via a deformation of the two
negative helicity gravitons. Our proof and the discussion in this
section validates the recursion relations used to construct the all
multiplicity ansatz. It would be highly desirable to show that the
formula proposed by Bedford et al. \cite{Bedford:2005yy} does indeed
satisfy the recursion relations. The formula is explicitly given by
\begin{equation}\label{BBST}
M_n(1^-,2^-,i_1^+,\ldots,i_{n-2}^+)=\frac{\langle
1,2\rangle^{6}[1,i_{n-2}]}{\langle 1,i_{n-2}
\rangle}G(i_1,i_2,i_3)\prod_{s=3}^{n-3}\frac{\langle
2|i_1+\ldots+i_{s-1}|i_s]}{\langle i_s,i_{s+1}\rangle \langle 2,
i_{s+1} \rangle}+\mathcal{P}(i_1,\ldots,i_{n-2})
\end{equation}
where ${\cal P}(i_1,\ldots,i_{n-2})$ indicates a sum over all
permutations of $(i_1,\ldots,i_{n-2})$ and
\begin{equation}\label{G}
G(i_1,i_2,i_3)=\frac{1}{2}\left( \frac{[i_1,i_2]}{\langle
2,i_1\rangle \langle 2, i_2\rangle \langle i_1,i_2\rangle \langle
i_2,i_3 \rangle \langle i_1,i_3 \rangle}\right).
\end{equation}

It is also interesting to show why the case
$(h_{i},h_{j})\,=\,(-,+)$ does not lead to recursion relations.
Using the Ward identity (\ref{wardi}) once again we have
\begin{equation}\label{WardMHV3}
M_{n}^{\mbox{\tiny MHV}}(i^{-} (z),j^{+}(z),l^{+},
 m^{-})\:=\:
 \left(\frac{\langle \lambda^{(i)}(z),\lambda^{(m)}\rangle}
 {\langle \lambda^{(i)}(z),\lambda^{(j)}\rangle} \right)^8
M_{n}^{\mbox{\tiny MHV}}( i^{-}(z),j^{-}(z),l^{+},
 m^{+})
\end{equation}
The amplitude on the right hand-side vanishes as $z$ goes to
infinity. However, $\langle
\lambda^{(i)}(z),\lambda^{(m)}\rangle^{8}$ contributes with a factor
of $z^{8}$ while $\langle \lambda^{(i)}(z),\lambda^{(j)}\rangle$ is
$z$ independent. Either using BGK (together with (\ref{WardMHV2}))
or directly (\ref{BBST}), one can show that $M_{n}^{\mbox{\tiny
MHV}}(i^{-}(z),j^{-}(z),l^{+},
 m^{+})$ goes like $1/z^2$, therefore the amplitude with
$(h_{i},h_{j})\,=\,(-,+)$ behaves as $z^6$ at infinity.

\section{Conclusions And Further Directions}

In this paper we have proven that tree level gravity amplitudes in
General Relativity are very special. Contrary to what can be called
a naive power counting of the behavior of individual Feynman
diagrams, full amplitudes actually vanish when momenta are taken to
infinity along some complex direction. The naive power counting
gives that the amplitudes diverge. This miraculous property implies
that tree amplitudes of gravitons satisfy a special kind of
recursion relations. One in which an amplitude is given as a sum of
terms containing the product of two physical on-shell amplitudes
where the momenta of only two gravitons have been complexified.
These recursion relations, originally discovered in
\cite{Britto:2004ap} in gauge theory, were proven using the power of
complex analysis in \cite{Britto:2005fq}. The BCFW construction
opened up the possibility for using complex analysis in many other
situations. There are only two major difficulties when applying the
BCFW construction to a general field theory at any order in
perturbation theory. One of them is that complete control of the
singularity structure of the amplitude is required. At tree-level
this means poles but at the loop level one can also have branch
cuts. The other one is to have a good control on the behavior at
infinity. In the case of gravity amplitudes this had been the
stumbling block. The way we overcome this obstacle was by
constructing auxiliary recursion relations. These were obtained by
exploiting as many polarization tensors as possible in other to tame
the divergent behavior of vertices in individual Feynman diagrams
while still keeping the linear behavior of propagators. In a sense,
the deformation we introduced is the ``maximal" choice.

This procedure seems quite general and it would be very interesting
to classify field theories according to whether their amplitudes
vanish or not at infinity under this maximal deformation.

\centerline{\bf Note Added (November 23, 2008):}

In the original version of this paper we made a wrong statement in the
introduction. We claimed that the deformation of \cite{Bjerrum-Bohr:2005jr} follows from the
basic BCFW deformation proven in this paper. However, the precise
statement made in \cite{Bjerrum-Bohr:2005jr} is that a proof of the basic BCFW deformation
would provide evidence for their deformation. In fact, it has recently
been proven in \cite{Bianchi:2008pu} by using direct numerical analysis that
the deformation of \cite{Bjerrum-Bohr:2005jr} generically fails to vanish at infinity for
$n\ge 12$ which means that the validity of the basic BCFW deformation is a
necessary but not sufficient condition for the validity of the non-maximal
deformations. The precise result found in \cite{Bianchi:2008pu} is for NMHV amplitudes and states that under a Risager deformation of the three negative helicity gravitons the large $z$ behavior is given by $z^{n-12}$. Given that this result might have important consequences we applied some of the techniques in this paper to construct an analytic proof. We provide the analytic proof of such behavior in appendix \ref{zn12}.

\begin{acknowledgments}

PB would like to thank Perimeter Institute for hospitality during a
visit where part of this research was done. PB is also grateful to
Fiorenzo Bastianelli for hospitality at the Department of Physics,
University of Bologna. The research of CBV and FC at Perimeter
Institute for Theoretical Physics is supported in part by the
Government of Canada through NSERC and by the Province of Ontario
through MRI. CBV also acknowledges support from NSERC Canadian
Graduate Scholarship.

\end{acknowledgments}

\appendix

\section{Proof Of Auxiliary Recursion Relations}

In the main part of the paper we used certain auxiliary recursion
relations to prove that $M_n(z)$ vanishes as $z$ is taken to
infinity under the BCFW deformation. It is therefore very important
to establish the validity of the auxiliary recursion relations.

Consider the case when then number of positive helicity gravitons is
larger or equal than the number of negative helicity ones, i.e,
$p\geq m$. The case when $m\geq p$ is completely analogous. Let us
start by constructing a rational function $M_n(w)$ of a complex
variable $w$ via the deformation (\ref{csd}), i.e,
\begin{equation}
\label{csdape} \tilde\lambda^{(j)}(w) = \tilde\lambda^{(j)} - w
\sum_{s\in \{k^+\}}\tilde\lambda^{(s)}, \qquad \lambda^{(k)}(w) =
\lambda^{(k)} + w\lambda^{(j)}, \qquad \forall k\in \{ k^+ \}
\end{equation}
where $j$ is a negative helicity graviton and $\{k^+ \}$ is the set
of all positive helicity gravitons in $M_{n}$.

The claim is that $M_n(w)$ vanishes as $w$ is taken to infinity and
its only singularities are simple poles at finite values of $w$.

\subsection{Vanishing Of $M_n(w)$ At Infinity}

Let us prove that $M_n(w)$ vanishes as $w\to \infty$. Consider the
leading Feynman diagram that contributes to $M_n(w)$. Such a diagram
has $n-2$ cubic vertices each contributing a factor of $w^2$. It
also has $p+1$ polarization tensors that depend on $w$ and give
$1/w^2$ each. Finally, we claim that all $n-3$ propagators that can
possibly depend on $w$ actually do giving each a contribution of
$1/w$. Putting all contributions together we find that the leading
Feynman diagrams go like $1/w^{p-m+3}$. Therefore, if $p\geq m$ then
$M_n(w)\to 0$ as $w\to \infty$.

We are only left to prove that $n-3$ propagators depend on $w$. A
similar statement has to be proven in section III.D. The proof there
is more involved since it requires the study of many cases. The
discussion that follows can be thought of as a warm up for that in
section III.D.

Consider a given Feynman diagram. A propagator naturally divides the
diagram into two sub-diagrams. Let us denote them by ${\cal I}$ and
${\cal J}$. Without loss of generality, we can always take the
$j^{\rm th}$ graviton to be in ${\cal J}$. Let us denote the set of
positive helicity gravitons in ${\cal I}$ by ${\cal I}^+$.

The propagator under consideration has the form $1/P^2_{\cal I}(w)$
with
\begin{equation}
\label{propi} P^2_{\cal I}(w) = P^2_{\cal I} - w\sum_{k\in {\cal
I}^+}\langle j|P_{\cal I}|k]
\end{equation}
where $P_{\cal I} = P_{\cal I}(0)$.

The only way the $w$ dependence can drop out of the propagator is
that $\sum_{k\in {\cal I}^+}\langle j|P_{\cal I}|k] =0$.

Since the $j^{\rm th}$ graviton belongs to ${\cal J}$, the condition
$\sum_{k\in {\cal I}^+}\langle j|P_{\cal I}|k] =0$ can only be
satisfied if the vector $\sum_{k\in {\cal I}^+}P_{{\cal I}\; a\dot
a}\tilde\lambda^{(k)\; \dot a}$ vanishes. To see this note that
there must be at least two gravitons in $\cal J$, one of them $j$.
Therefore we can use momentum conservation to determine the other
one in terms of the other $n-1$ gravitons. This allows us to
consider all the remaining $n-1$ gravitons as independent. In
particular, the $j^{\rm th}$ graviton is independent from the ones
in $\cal I$.

Our goal is then to prove that the combination $P_{{\cal I} \; a\dot
a}(\sum_{k\in {\cal I}^+}\tilde\lambda^{(k)}_{\dot a})$ cannot
vanish for generic choice of momenta and polarization tensors.

Consider first the case when the set ${\cal I}^+$ has only one
element, say the $s^{\rm th}$ graviton. Then the vanishing of
$P_{{\cal I} \; a\dot a}\tilde\lambda^{(s)\; \dot a}$ implies that
of $\sum_{k\in {\cal I}}s_{k,s}$, where $s_{k,s} = (p_k+p_s)^2$.
Since $\cal I$ must have at least two gravitons, the vanishing of
$\sum_{k\in {\cal I}}s_{k,s}$ is a constraint on the kinematical
invariants which is not satisfied for generic momenta.

Consider the case when ${\cal I}^+$ has at least two elements. Let
one of them be the $s^{\rm th}$ graviton. Since our starting point
is a physical on-shell amplitude, the dependence of the amplitude on
the $s^{\rm th}$ graviton can only be through its polarization
tensor and its momentum vector,
\begin{equation}
\epsilon^{+\; (s)}_{a\dot a, b \dot b} =
\frac{\mu_a\tilde\lambda^{(s)}_{\dot
a}\mu_b\tilde\lambda^{(s)}_{\dot b}}{\langle \mu,
\lambda^{(s)}\rangle^2}, \qquad p^{(s)}_{a\dot a} =
\lambda_a^{(s)}\tilde\lambda_{\dot a}^{(s)}.
\end{equation}
If we transform $\{ \lambda^{(s)}, \tilde\lambda^{(s)}\}$ into $\{
t^{-1}\lambda^{(s)}, t\tilde\lambda^{(s)}\}$ with $t^4=1$, i.e., $t$
is any $4^{\rm th}$ root of unity, then both $\epsilon^{+\;
(s)}_{a\dot a, b \dot b}$ and $p^{(s)}_{a\dot a}$ are invariant.
This means that any statement we make for $t=1$ must be true for the
other three possible values of $t$. In particular, it must be the
case that $P_{{\cal I} \; a\dot a}(\sum_{k\in {\cal I}^+ , \; k\neq
s}\tilde\lambda^{(k)}_{\dot a} + t\tilde\lambda^{(s)}_{\dot a})$
vanishes for all four values of $t$. Since $P_{{\cal I} \; a\dot a}$
does not depend on $t$ the only way to satisfy this condition is if
$P_{{\cal I}}\cdot p^{(s)} =0$. This is clearly a condition that is
not satisfied for generic momenta and therefore this possibility is
also excluded.

Finally, there is one more possibility to consider. If the set
${\cal I}^+$ is empty then the $w$ dependence drops out. Of course,
this is not a problem because if the set ${\cal I}^+$ is empty it
means that nothing on the subdiagram ${\cal I}$ depends on $w$,
including the cubic vertices. Therefore, neither propagators nor
cubic vertices contribute. One can then concentrate on the
subdiagram ${\cal J}$, but this subdiagram has less particles than
the total diagram and the same number of $w$-dependent polarization
tensor. Therefore these diagrams go to zero even faster than
diagrams where ${\cal I}^+$ is not empty.

\subsection{Location Of Poles And Final Form Of The Auxiliary Recursion Relations}

Having proven that $M_n(w)$ vanishes at infinity, we turn to the
question of the singularity structure. We claim that it has only
simple poles coming from propagators in Feynman diagrams. Again as
in section II where we discussed the BCFW deformation, one has that
the poles generated by the $w$ dependence in the polarization
tensors can be eliminated by a gauge choice. We pick the reference
spinor of each of the polarization tensors of the positive helicity
gravitons to be $\mu_{a} = \lambda_{a}^{(j)}$ and that of the
$j^{\rm th}$ helicity graviton to be $\tilde\mu_{\dot a} =
\sum_{k\in \{ k^+\} }\tilde\lambda^{(k)}$.

We have already given the structure of propagators in (\ref{propi})
from where we can immediately read off the location of the poles to
be
\begin{equation}
w_{\cal I} = \frac{P^2_{\cal I}(0)}{\sum_{k\in {\cal I}^+}\langle
j|P_{\cal I}(0)|k]}.
\end{equation}

Finally, we need the fact that a rational function that vanishes at
infinity and only has simple poles can be written as $M_n(w) =
\sum_{\alpha} c_{\alpha}/(w-w_{\alpha})$ where the sum is over the
poles and $c_{\alpha}$ are the residues. The residues in this case
can be determined from factorization limits since all poles come
from physical propagators.

Collecting all results we arrive at the final form of the auxiliary
recursion relation used in the text (\ref{RR}):
\begin{equation}\label{RRapi}
\begin{split}
M_{n}&(\{ r^-\}, \{ k^+\})\:=\\
&=\:\sum_{\mathcal{I}}\sum_{h=\pm}
M_{\mathcal{I}}\left(\left\{r_{\mathcal{I}}^{-}\right\},
 \left\{k_{\mathcal{I}}^{+}(w_{\mathcal{I}})\right\},
 -P_{\mathcal{I}}^{h}(w_{\mathcal{I}})\right)\frac{1}{P^{2}_{\mathcal{I}}}
M_{\mathcal{J}}\left(\left\{r_{\mathcal{J}}^{-}(w_{\cal I})\right\},
 \left\{k_{\mathcal{J}}^{+}(w_{\mathcal{I}})\right\},
 P_{\mathcal{I}}^{-h}(w_{\mathcal{I}})\right).
\end{split}
\end{equation}

\section{Large-$z$ behavior of NMHV gravity amplitudes under a three-particle deformation}\label{zn12}

\begin{figure}
 \begin{equation*}
  \begin{split}
  M_{n}\:&=\:\sum_{l\in\hat{\mathcal{I}}^{+}}{\raisebox{-1.18cm}{\scalebox{.43}{{\includegraphics{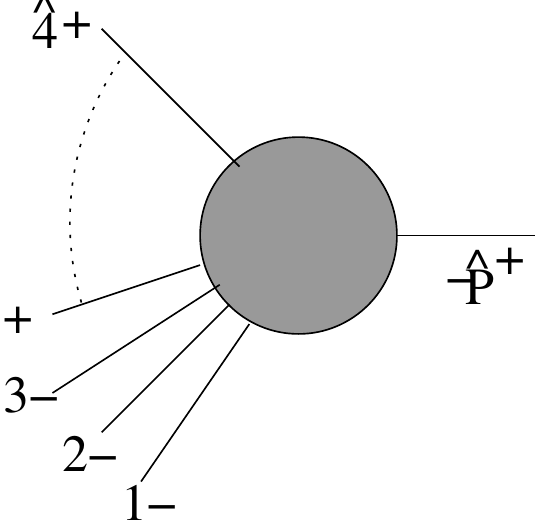}}}}}
           \:\frac{1}{P^{2}_{l5}}\:
	   {\raisebox{-1.10cm}{\scalebox{.43}{{\includegraphics{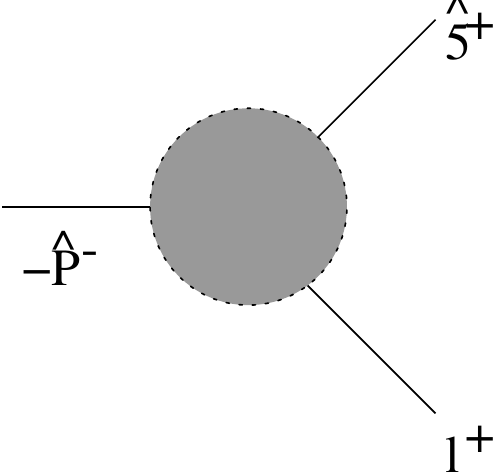}}}}}
	   \:+\:
           \sum_{i,j,k\in\{1,2,3\}}{\raisebox{-1.26cm}{\scalebox{.43}{{\includegraphics{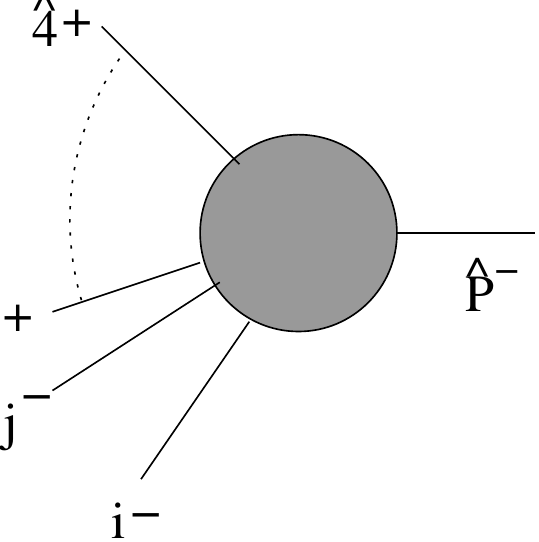}}}}}
	   \:\frac{1}{P^{2}_{k5}}\:
            {\raisebox{-1.01cm}{\scalebox{.43}{{\includegraphics{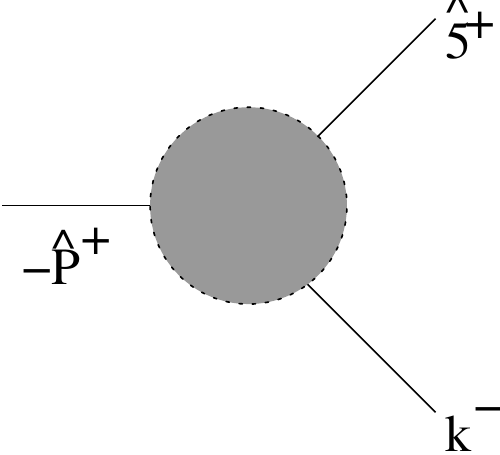}}}}}\\
	 &+\:\sum_{\hat{\mathcal{I}}^{+}}\sum_{i,j,k\in\{1,2,3\}}
	   {\raisebox{-1.25cm}{\scalebox{.43}{{\includegraphics{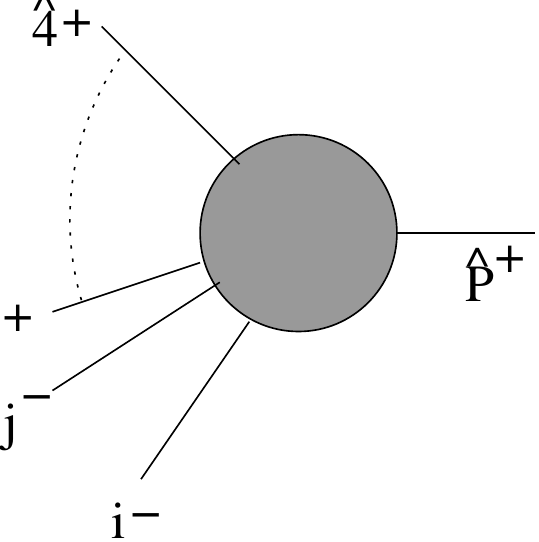}}}}}
           \:\frac{1}{P^{2}_{\hat{\mathcal{I}}^{+}}}\:
	   {\raisebox{-1.01cm}{\scalebox{.43}{{\includegraphics{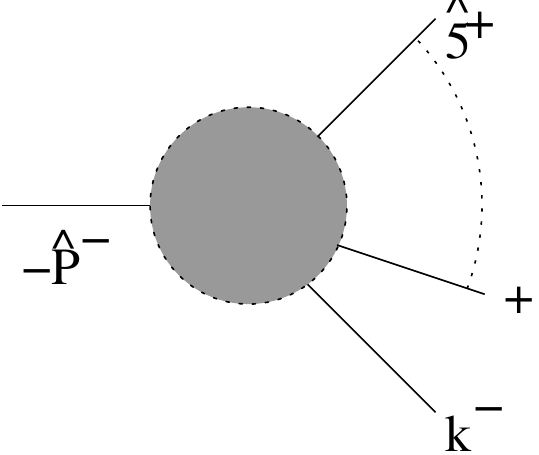}}}}}\\
	  &+\:
           \sum_{\hat{\mathcal{I}}^{+}}\sum_{i,j,k\in\{1,2,3\}}
	   {\raisebox{-1.25cm}{\scalebox{.43}{{\includegraphics{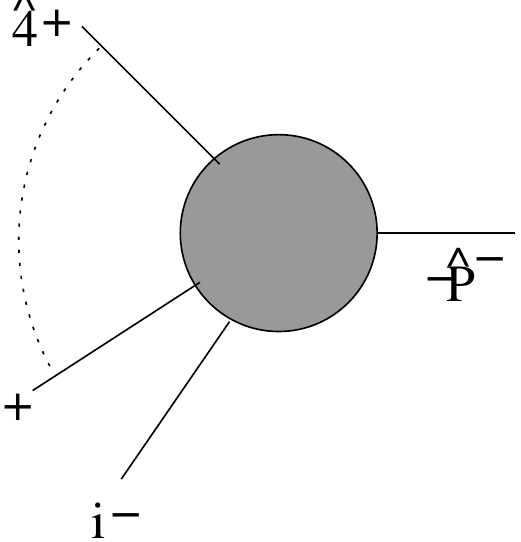}}}}}
           \:\frac{1}{P^{2}_{\hat{\mathcal{I}}^{+}}}\:
	   {\raisebox{-1.01cm}{\scalebox{.43}{{\includegraphics{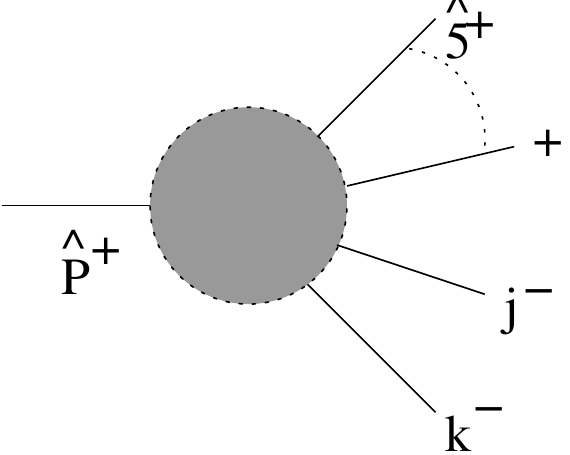}}}}}\\
	  &+\:\sum_{i,j,k\in\{1,2,3\}}{\raisebox{-1.26cm}{\scalebox{.43}{{\includegraphics{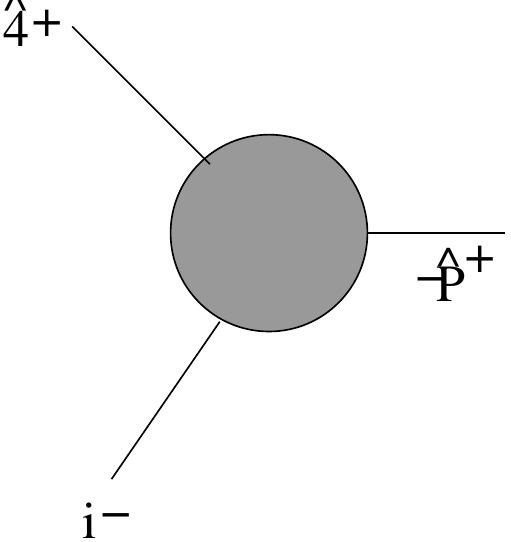}}}}}
	   \:\frac{1}{P^{2}_{i4}}\:
            {\raisebox{-1.01cm}{\scalebox{.43}{{\includegraphics{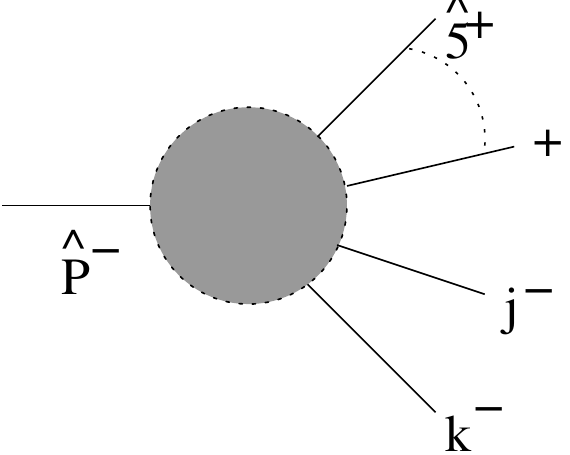}}}}}
	   \:+\:\sum_{l\in\hat{\mathcal{I}}^{+}}{\raisebox{-1.18cm}{\scalebox{.43}{{\includegraphics{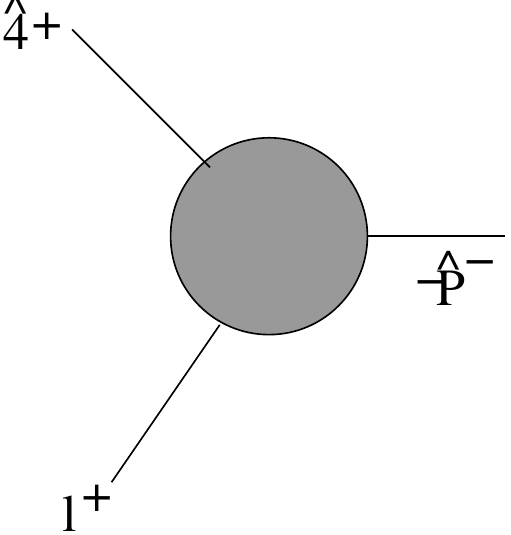}}}}}
           \:\frac{1}{P^{2}_{l4}}\:
	   {\raisebox{-1.35cm}{\scalebox{.43}{{\includegraphics{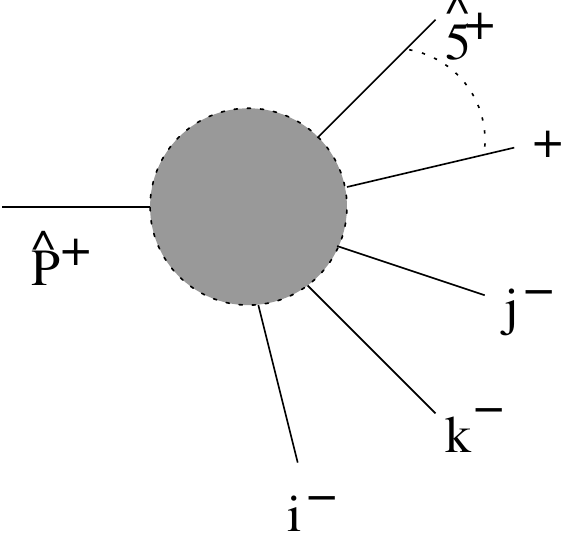}}}}}
  \end{split}
 \end{equation*}
 \caption{Representation of the NMHV gravity amplitude under a BCFW-like deformation of momenta of two positive
          helicity gravitons}\label{expanfig}
\end{figure}

In \cite{Bjerrum-Bohr:2005jr} the MHV-expansion for gravity scattering amplitude was discussed. The main
idea was to obtain it as a recursion relation: it was argued that the suitable one-parameter deformation
of the momentum space which could allow to explicitly obtain such an expansion can be introduced by shifting
three gravitons with same helicity as follows. For semplicity, let us consider an NMHV amplitude and let us
deform the momenta of three negative helicity gravitons $\{1,\,2,\,3\}$
\begin{equation}\label{3partdef}
 \tilde\lambda^{(i)}(z)\:=\:\tilde\lambda^{(i)}-z\langle j,\,k\rangle\tilde\eta, \qquad i,\,j,\,k\:=\:1,\,2,\,3,
\end{equation}
where $\tilde\eta$ is a reference spinor and the coefficients $\langle j,\,k\rangle$ are fixed by momentum
conservation (the indices $i,\,j,\,k$ assume the values $1,\,2,\,3$ cyclically). It was assumed that
the deformed amplitude was well-behaved as $z\rightarrow\infty$, which is the necessary and sufficient condition
to determine the amplitude only from its poles at finite points. Recently, it was shown in \cite{Bianchi:2008pu}
that actually this condition does not hold for an arbitrary number $n$ of external gravitons. More precisely,
it has been shown numerically that, under the deformation (\ref{3partdef}), the NMHV graviton amplitude vanishes as long as
the number of external gravitons is less than $12$:
\begin{equation}\label{Mzn12}
M_{n}^{\mbox{\tiny NMHV}}(z)\:\sim\:\frac{1}{z^{n-12}}.
\end{equation}
In this section we provide an analitycal proof of such behavior. The main idea of the following proof is to
consider a suitable representation for the NMHV graviton amplitude and then apply the three-particle deformation
(\ref{3partdef}). Such a representation is obtained through a BCFW-like deformation of two positive helicity
gravitons, which we label by $4$ and $5$. It is important to mention that the validity of such a deformation does not follow from the one proven in this paper where opposite helicity gravitons are deformed. Luckily, a proof that equal helicity gravitons can also be defomed was recently given by Arkani-Hamed and Kaplan in \cite{ArkaniHamed:2008yf}. 
As shown in Figure \ref{expanfig}, this particular deformation induces a diagrammatic
expansion for the amplitude with $6$ types of contributions.
\begin{equation}\label{expan}
 \begin{split}
  M_{n}\:&=\:\sum_{l\in\hat{\mathcal{I}}^{+}}
           M_{n-1}\left(\left\{r^{-}\right\},\left\{\hat{\mathcal{I}}^{+}_{l}\right\},\hat{4}^{+},\hat{P}_{l5}^{+}\right)
            \frac{1}{P_{l5}^{2}}
           M_{3}\left(-\hat{P}_{l5}^{-},l^{+},\hat{5}^{+}\right)+\\
	 &+\:\sum_{i,j,k\in\{1,2,3\}}
	 M_{n-1}\left(i^{-},j^{-},\left\{\hat{\mathcal{I}}^{+}\right\},\hat{4}^{+},\hat{P}_{k5}^{-}\right)
            \frac{1}{P_{k5}^{2}}
           M_{3}\left(-\hat{P}_{k5}^{+},k^{-},\hat{5}^{+}\right)+\\
         &+\:\sum_{\hat{\mathcal{I}}^{+}}\sum_{i,j,k\in\{1,2,3\}}
	 M_{\hat{\mathcal{J}}^{+}}\left(i^{-},j^{-},\left\{\hat{\mathcal{J}}^{+}\right\},\hat{4}^{+},\hat{P}_{\hat{\mathcal{I}}}^{+}\right)
	   \frac{1}{P_{\hat{\mathcal{I}}^{+}}^{2}}
	 M_{\hat{\mathcal{I}}^{+}}\left(-\hat{P}_{\hat{\mathcal{I}}}^{-},\left\{\hat{\mathcal{I}}^{+}\right\},k^{-},\hat{5}^{+}\right)+\\
	 &+\:\sum_{\hat{\mathcal{I}}^{+}}\sum_{i,j,k\in\{1,2,3\}}
	 M_{\hat{\mathcal{I}}^{+}}\left(i^{-},\left\{\hat{\mathcal{I}}^{+}\right\},\hat{4}^{+},-\hat{P}_{\hat{\mathcal{I}}}^{-}\right)
	   \frac{1}{P_{\hat{\mathcal{I}}^{+}}^{2}}
	 M_{\hat{\mathcal{J}}^{+}}\left(\hat{P}_{\hat{\mathcal{I}}}^{+},\left\{\hat{\mathcal{J}}^{+}\right\},j^{-},k^{-},\hat{5}^{+}\right)+\\
	 &+\:\sum_{i,j,k\in\{1,2,3\}}
	 M_{3}\left(i^{-},\hat{4}^{+},-\hat{P}_{i4}^{+}\right)
            \frac{1}{P_{i4}^{2}}
         M_{n-1}\left(\hat{P}_{i4}^{-},j^{-},k^{-},\left\{\hat{\mathcal{J}}^{+}\right\}\hat{5}^{+}\right)+\\
         &+\:\sum_{l\in\hat{\mathcal{I}}^{+}}
	   M_{3}\left(l^{+},\hat{4}^{+},-\hat{P}_{l4}^{-}\right)
            \frac{1}{P_{l4}^{2}}
           M_{n-1}\left(\left\{r^{-}\right\},\left\{\hat{\mathcal{I}}^{+}_{l}\right\},\hat{5}^{+},\hat{P}_{l4}^{+}\right),
  \end{split}
\end{equation}
where $\hat{\mathcal{I}}^{+}\,\equiv\mathcal{I}^{+}\setminus\left\{4^{+},5^{+}\right\}$ and
$\hat{\mathcal{I}}^{+}_{l}\,\equiv\hat{\mathcal{I}}^{+}\setminus\left\{l^{+}\right\}$.
Let us write explicitly down the deformation we are using
\begin{equation}\label{2plus}
 \tilde\lambda^{(5)}(w)\:=\:\tilde\lambda^{(5)}-w\tilde\lambda^{(4)},\qquad
 \lambda^{(4)}(w)\:=\:\lambda^{(4)}+w\lambda^{(5)}.
\end{equation}
It is possible to see immediately that the first two lines in (\ref{expan}) vanish, given that the three-particle amplitudes
present are zero: those three-particle amplitudes are anti-holomorphic and the anti-holomorphic spinors involved
are proportional to each other, so that $M_{3}\,=\,0$.

Let us now consider the terms in the last line of (\ref{expan}). The poles in this channel are located at
$w_{l4}\,=\,\langle l,4\rangle/\langle 5,l\rangle$, and the deformed spinors evaluated at these poles are given by
\begin{equation}\label{spin6}
 \hat{\tilde\lambda}^{(5)}\:=\:\tilde\lambda^{(5)}-\frac{\langle l,4\rangle}{\langle5,l\rangle}\tilde\lambda^{(4)},\qquad
 \hat{\lambda}^{(4)}\:=\:\frac{\langle5,4\rangle}{\langle5,l\rangle}\lambda^{(l)}.
\end{equation}
Similarly, the spinors related to on-shell ``internal'' momentum $P_{l4}$ can be written as
\begin{equation}\label{Pl4}
 \lambda^{(P)}\:=\:\lambda^{(l)},\qquad
 \tilde{\lambda}^{(P)}\:=\:\tilde{\lambda}^{(l)}+\frac{\langle5,4\rangle}{\langle5,l\rangle}\tilde{\lambda}^{(4)}.
\end{equation}
It is straightforward then to see that, under the deformation (\ref{3partdef}) both the propagator $1/P_{l4}^{2}$ and
the three-particle amplitude are $z$-independent, while the $(n-1)$-particle amplitude gets deformed in the same way
of the original amplitude. These terms can therefore be treated by induction.

We repeat the same strategy for all the other terms in the BCFW-representation, starting with the fifth line in
(\ref{expan}). In this channel the poles are located at $w_{i4}\,=\,\langle i,4\rangle/\langle5,i\rangle$,
and the deformed spinors evaluated at these poles writes
\begin{equation}\label{spin5}
 \hat{\tilde\lambda}^{(5)}\:=\:\tilde\lambda^{(5)}-\frac{\langle i,4\rangle}{\langle5,i\rangle}\tilde\lambda^{(4)},\qquad
 \hat{\lambda}^{(4)}\:=\:\frac{\langle5,4\rangle}{\langle5,i\rangle}\lambda^{(i)},
\end{equation}
with
\begin{equation}\label{Pi4}
 \lambda^{(P)}\:=\:\lambda^{(i)},\qquad
 \tilde{\lambda}^{(P)}\:=\:\tilde{\lambda}^{(i)}+\frac{\langle5,4\rangle}{\langle5,i\rangle}\tilde{\lambda}^{(4)}.
\end{equation}
Contrarly to the previous case, here the three-particle deformation (\ref{3partdef}) induces a non-trivial
$z$-dependence in (\ref{Pi4}) given that there is an explicit dependence on one of the deformed
spinors (namely the one labeled by $i$). It is interesting to notice the way they get deformed:
\begin{equation}\label{Pi4def}
 \lambda^{(P)}(z)\:=\:\lambda^{(P)},\qquad
 \tilde{\lambda}^{(P)}(z)\:=\:\tilde{\lambda}^{(P)}-z\langle j,k\rangle\tilde{\eta},
\end{equation}
which implies that for the three-particle amplitude we have:
\begin{equation}\label{spin5amp}
  M_{3}\left(i^{-},\hat{4}^{+},-\hat{P}_{i4}^{+}\right)\:=\:\left(\frac{[4,P(z)]^3}{[P(z),i(z)][i(z),4]}\right)^{2}
  \:=\:\left(\frac{\langle5,i\rangle}{\langle5,4\rangle}\left([4,i]-z\langle j,k\rangle[4,\tilde\eta]\right)\right)^2\:\sim\:z^2,
\end{equation}
while the propagator $1/P_{i4}^2(z)$ behaves as $1/z$ and $(n-1)$-particle amplitude again is deformed according to a shift
of type (\ref{3partdef}). This means that the product of the three-particle amplitude and the propagator behaves as $\sim z$ for large $z$.

Note that this is very different from the behavior in Yang-Mills where the three-particle amplitude is the square root of that in gravity and its large $z$ behavior is canceled by that of the propagator. This means that in Yang-Mills the same product goes like $~1$ for large $z$.

In order to conclude the argument, we simply have to notice that the amplitude multiplying the ${\cal O}(z)$ term in gravity (and ${\cal O}(1)$ term in Yang-Mills) is a $(n-1)$-particle amplitude under the three-particle deformation (\ref{3partdef}). This means that if the $(n-1)$-particle amplitude behaves as $z^{a_{n-1}}$ at infinity then the $n$-particle amplitude behaves as $z^{a_{n-1}+1}$ in gravity (and as $z^{a_{n-1}}$ in Yang-Mills). All we need is to find out what $a_n$ is and this can be done by studying explicitly low point amplitudes.
Using the explicit form of the amplitude, we have checked that a five particle amplitude behaves as $z^{-7}$. Therefore $a_5 =-7$. Using that $a_n = a_{n-1}+1$ we find $a_n = n-12$ which is what we wanted to prove. In Yang-Mills one finds that $a_n=a_{n-1}$ and therefore the amplitude vanishes at infinity if it does for any $n$. It is easy to check that this is indeed the case for $n=5$. 

%
%

\end{document}